\documentclass[fleqn,usenatbib]{mnras}

\usepackage[T1]{fontenc}

\DeclareRobustCommand{\VAN}[3]{#2}
\let\VANthebibliography\thebibliography
\def\thebibliography{\DeclareRobustCommand{\VAN}[3]{##3}\VANthebibliography}

\usepackage{graphicx}	
\usepackage{amsmath}	
\usepackage{amssymb}	

\usepackage{multirow}
\usepackage{subcaption}

\usepackage{newtxtext,newtxmath}

\newcommand{\lya}{Ly$\alpha$ }

\defcitealias{2019MNRAS.487.1047M}{M19}
\defcitealias{2020MNRAS.499.1640M}{MM20}

\title[X-ray impact on \ion{H}{I} reionization fossils]{The long-lasting effect of X-ray preheating in the post-reionization intergalactic medium}

\author[Montero-Camacho, Zhang \& Mao]{
Paulo Montero-Camacho,$^{1,2}$\thanks{E-mail: pmontero@pcl.ac.cn (PMC)} Yao Zhang$^{2}$ and Yi Mao$^{2}$\thanks{E-mail: ymao@tsinghua.edu.cn (YM)}  
\\
$^{1}$Department of Mathematics and Theory, Peng Cheng Laboratory, Shenzhen, Guangdong 518066, China\\
$^{2}$Department of Astronomy, Tsinghua University, Beijing 100084, China\\
}

\date{Accepted XXX. Received YYY; in original form ZZZ}

\pubyear{2023}

\begin{document}
\label{firstpage}
\pagerange{\pageref{firstpage}--\pageref{lastpage}}
\maketitle

\begin{abstract}
X-ray photons can penetrate deep into the intergalactic medium (IGM), leading to preheating of the IGM prior to cosmic reionization. X-ray preheating wipes out some of the small-scale structures that would otherwise be present prior to the passage of an ionization front. Accurate modeling of the small-scale structure is vital to the post-reionization IGM since the small-scale structure is ultimately the dominant source of long-lasting relics from hydrogen reionization. However, the precise impact of X-ray preheating in the fossils from hydrogen reionization is highly uncertain. In this work, we explore and establish for the first time, the long-lasting impact of X-ray preheating in the post-reionization IGM via hydrodynamic simulations with high-mass resolution. We find that the addition of X-ray preheating astrophysics leads to an overall lesser impact of the effect of inhomogeneous reionization in the \lya forest -- depending on specific X-ray prescription -- at low redshifts ($z \sim 2$) with respect to a model with no X-ray preheating. However, at high redshifts ($z \sim 4$), our results indicate a strengthening of the relics of reionization in the \lya forest because the IGM becomes more transparent compared to the scenario with no preheating. Thus, the absence of X-ray preheating in \lya modeling can lead to a biased inference of cosmological parameters. Nevertheless, optimistically, the inclusion of X-ray preheating emerges as a promising novel avenue to probe the astrophysics of cosmic dawn.
\end{abstract}




\section{Introduction}
The Universe is currently believed to transition from primarily neutral to mainly ionized around $z \sim 7.7$ \citep{2018arXiv180706209P}. This phase transition, known as cosmic reionization, occurs via the passage of ionization fronts through the intergalactic medium (IGM) resulting in a violently perturbed IGM that gets heated up to $\sim 10^4 $ K \citep{1994MNRAS.266..343M,2009ApJ...701...94F,2021ApJ...906..124Z}. The heating of the IGM is primarily driven by ultraviolet (UV) photons that originate from the first stars and galaxies, and thus reionization happens in a patchy fashion where denser regions reionize first due to a larger abundance of UV sources \citep[see e.g.][for a review]{2016ARA&A..54..313M}. Although the physics of reionization are -- to a certain extent -- understood, finer details like the timeline of the reionization process \citep[e.g.][]{2021ApJ...919..120M,2022MNRAS.514...55B,2022MNRAS.517.1264L,2022arXiv221112613J} and the nature and evolution of the ionizing sources \citep{2021ApJ...917...58L,2022arXiv220711278K,2022MNRAS.512.4909G} are currently the focus of ongoing efforts. 

Furthermore, the violent injection of energy into the IGM during reionization coupled with the inhomogeneous nature of reionization leads to relics in the post-reionization IGM that can survive to low redshifts ($z<4$) \citep{2018MNRAS.474.2173H}. These imprints of reionization, which affect the evolution of the post-reionization IGM \citep{2008ApJ...689L..81T,2019MNRAS.490.3177W,2019MNRAS.486.4075O}, can be categorized into two types -- the ``void response'' and the ``halo response''. The void response is responsible for the imprints of reionization that affect the \lya forest at low redshifts ($z \lesssim 4$) (\citealt{2019MNRAS.487.1047M}, hereafter referred to as \citetalias{2019MNRAS.487.1047M}). The reason the (mini) voids play a crucial role is that they not only get ultraviolet heated by the ionization sources but also experience shock heating, and accompanying compression to mean density, from shocks that originate in the surrounding denser regions. Overall, this results in a long-lasting impact since they can get pushed to a higher adiabat than the denser gas and can reach temperatures of $\sim 3 \times 10^4$ K \citep{2018MNRAS.474.2173H}. On the other hand, we define the halo response to encompass the effects that affect the denser regions. Gas in these regions can likely recover from the reionization ``energetic kick'' by $z \sim 3.5 - 4$ \citep{2022arXiv221002385L}. The halo response, i.e.\ the imprints left by the passage of ionization fronts on shallow potential wells, not only destroy small-scale baryonic structure ($\sim 10^6$ M$_\odot$) \citep{2004MNRAS.348..753S} but also suppress the growth of structure for cosmological time scales,  eventually reaching a filtering scale of $\sim 100$ kpc after hundreds of millions of years \citep{2020ApJ...898..149D}. After cosmic reionization, the thermal evolution of the IGM was believed to follow a stringent power law, the so-called temperature-density relation \citep{1997MNRAS.292...27H}, which arises as a consequence of the scaling of the recombination rate with temperature and it is facilitated by both adiabatic expansion and Compton cooling \citep{2016MNRAS.456...47M}. Although both types of reionization relics affect the relaxation/recovery of the temperature-density relation, only the void response leads to signatures in the \lya forest at low redshifts ($z \leq 4)$. On the other hand, the evolution of the filtering of masses, i.e.\ how many baryons can sit on a given small halo host, is greatly impacted by the halo response to the passage of ionization fronts \citep{2022MNRAS.513..117L} and it is vital to compute the impact of inhomogeneous reionization in the post-reionization \ion{H}{I} 21 cm intensity mapping. Note that both void and halo responses are somewhat artificial separations made here for an easier understanding. Besides, both effects need to be considered while simultaneously accounting for the inhomogeneous nature of reionization for an accurate model of the post-reionization IGM.

These imprints from hydrogen reionization leave strong ripples that cosmological probes of the post-reionization IGM can recover. For instance, for the \lya forest, i.e.\ the absorption features due to neutral hydrogen absorption of Lyman-$\alpha$ radiation emitted from distant quasars, its power spectra have been shown to suffer an enhancement at large scales due to the fossils from \ion{H}{I} reionization (\citetalias{2019MNRAS.487.1047M}; \citealt{2022MNRAS.509.6119M,2022MNRAS.tmp.3519P})\footnote{However, \citet{2022ApJ...928..174M} found no enhancement at large scales; nevertheless, relics from cosmic reionization in the post-reionization era were still present. The reason for this disagreement is still under investigation.}. In addition, the line intensity mapping (LIM) of neutral hydrogen 21~cm hyperfine transition also suffers a similar effect, which is driven by the second type of relic, i.e.\ the hydrodynamic response \citep{2022arXiv221002385L}.

Naturally, this novel broadband signal, which originates during the epoch of reionization, depends on the astrophysics that governs cosmic reionization (\citealt{2020MNRAS.499.1640M}, hereafter referred to as \citetalias{2020MNRAS.499.1640M}). The ability to learn about the timeline of reionization \citep{2021MNRAS.508.1262M} or to separate cosmology from reionization astrophysics \citep{2023MNRAS.520.4853M} will depend on a clear holistic understanding of the reionization process and on accurate handling of other astrophysical and instrumental systematics that will be present in both 21~cm LIM and \lya forest. Here in this paper, we will focus on one piece of the puzzle, namely the impact of X-ray preheating in the post-reionization IGM.

During a period preceding cosmic reionization, X-rays heat the IGM up and raise the gas temperature from below to above the cosmic microwave background (CMB) temperature, leading to a transition of observing this epoch of heating from absorption to emission through the 21~cm line relative to the CMB. X-rays have small cross-sections for interactions with atoms. Hence, they penetrate deep into the neutral IGM and subsequently preheat the IGM through photoelectric absorption. However, the impact of X-ray preheating in the high-redshift IGM ($z \sim 15$) is highly uncertain due to a lack of observational constraints. Thus, theoretical models of the IGM often consider medium IGM temperatures ranging from tens to hundreds of Kelvins \citep[see Figure 4 of ][]{2017MNRAS.468.3785R}. Furthermore, other sources of uncertainty are the extrapolation of low-redshift X-ray luminosity functions to high redshifts \citep{2022ApJ...930..135L}, uncertainty regarding the abundance, clustering, and nature of the heating sources, including X-ray binaries \citep[see, e.g.][]{2014ApJ...791..110X,2017ApJ...840...39M,2022MNRAS.513.5097K,2023arXiv230303435S}, high-$z$ quasars and active galactic nuclei \citep{2011MNRAS.417.2264V,2015ApJ...813L...8M,2019MNRAS.487.1101R}\footnote{Nevertheless,  \cite{2018MNRAS.476.1174E} claimed that, based on their seeding prescription, nuclear black holes are too scarce and thus do not contribute significantly to heating or ionization at $z>10$.}, cosmic rays from supernovae in Population III stars \citep{2023arXiv230407201G}, and hot interstellar medium emission \citep{2014MNRAS.443..678P,2021ApJ...912..143M}.

As a result of X-ray preheating, some structures at small scales ``puff up'' before the passage of an ionization front. Therefore, the relaxation into the usual temperature-density relation of the IGM should occur faster \citep{2018MNRAS.474.2173H}, which would lead to an overall suppression of the large-scale enhancement in both the \lya forest and on the 21~cm LIM imprint in the post-reionization era. In addition, the presence of an anisotropic radiation field due to the sources of X-ray photons impacts the high-redshift IGM in several ways. For instance, X-rays should first increase the 21~cm power spectrum on large scales \citep{2007MNRAS.376.1680P,2016MNRAS.458.2710E}, followed by an overall reduction of the strength of 21~cm fluctuations due to X-rays wiping off temperature fluctuations in small scales. Moreover, the presence of such energetic sources is expected to leave clear imprints on the 21~cm bispectrum \citep{2019MNRAS.482.2653W,2021ApJ...912..143M} and to significantly increase non-Gaussianity \citep{2017MNRAS.468.3785R, 2019MNRAS.487.1101R}. Furthermore, X-ray preheating yields a sizable contribution to the early stages of reionization, which can lead to an earlier, albeit slightly extended, epoch of reionization and a more uniform \ion{H}{II} morphology \citep{2013MNRAS.431..621M}. 

Recently, \cite{2022arXiv221004912T} has placed constraints on the X-ray luminosity and on the effect of X-ray preheating in the cosmic dawn, disfavoring the extrapolation of the local relationship between soft X-ray luminosity and star formation \citep{2012MNRAS.419.2095M} to high redshifts. On the other hand, their results are consistent with X-rays being produced by a population of low-metallicity high-mass X-ray binaries \citep{2013ApJ...764...41F}. Moreover, \cite{2022arXiv221004912T} finds that the intergalactic medium must have been heated above the CMB temperature of $\sim 31\,{\rm K}$ as early as $z = 10.4$. 

In principle, preheating of the IGM could leave observable signatures in the post-reionization era through two main effects. First, X-rays will wipe out some of the small-scale structures that would otherwise be present. Second, X-ray preheating sets up the initial conditions for cosmic reionization since X-ray photons heat up the neutral gas before the UV ionizing sources, thus leading to a contribution to the global ionized hydrogen fraction prior to the start of proper cosmic reionization ($z \sim 12$). Regarding the latter effect, \citetalias{2020MNRAS.499.1640M} demonstrated that the acceleration or deceleration of the initial stages of reionization, i.e. increase or decrease of $\overline{x}_{\rm HII} (z \sim 12)$, has a minimal impact on the long-lasting relics from reionization in the IGM (as long as non-extreme scenarios of X-ray preheating are considered). In this work, we focus on the former effect, i.e.\ we investigate the impact of the ``puffing-up'' of structures during the early stages of the reionization process. We highlight that assessing the dependence of the impact of inhmogeneous reionization on cosmic dawn astrophysics is a vital step toward future Bayesian frameworks for the inference of unbiased cosmological parameters from cosmological probes of the post-reionization era \citep{2023MNRAS.520.4853M}.  

The rest of this paper is organized as follows. We introduce the different models for X-ray preheating of the IGM considered throughout this work in \S\ref{sec:preheat} and describe our simulations in \S\ref{sec:meth}. In \S\ref{sec:igm}, we establish the impact of preheating on the IGM. In particular, we quantify the expected impact on the 3D and 1D \lya power spectrum in \S\ref{ssec:3d} and \S\ref{ssec:1d}, respectively. Note that in order to facilitate the exploration of several different X-ray preheating models, the calculations in \S\ref{sec:igm} sacrifice accuracy due to the use of small box lengths. Thus, we make a separate analysis with larger boxes but with only a single X-ray preheating model in \S\ref{sec:II}. We summarize our findings in \S\ref{sec:conc}. We show the impact of preheating on the IGM temperature-density relation in Appendix \ref{app:hemd}. Some intermediate, tabulated results are left to Appendix~\ref{app:I}.

Throughout this work, we assume the following fiducial cosmology: $h=0.6774$, $\Omega_{\rm c}h^2=0.1417$, $\Omega_{\rm b}h^2=0.02230$, $A_{\rm s}=2.142\times 10^{-9}$, $n_{\rm s}=0.9667$. Our chosen cosmology is close to Planck 2015 ``TT + TE + EE + lowP + lensing + ext'' result \citep{2016A&A...594A..13P}. Our fiducial model has no X-ray preheating.

\section{Preheating of the IGM}
\label{sec:preheat}
During the epoch of X-ray heating, X-ray photons are expected to raise the temperature of the IGM to higher levels than the CMB temperature, thereby leading to an observational signature in the 21~cm signal. Although the precise degree of heating and its redshift evolution is uncertain, it is nonetheless vital to understand X-ray preheating since it is degenerate with reionization astrophysics \citep{2016MNRAS.458.2710E}. However, the reionization relics in the post-reionization IGM were found to be insensitive to X-ray preheating \citepalias{2020MNRAS.499.1640M}, thus allowing for a possible window for constraining the astrophysics of reionization without worrying about these degeneracies \citep{2021MNRAS.508.1262M}. Here we revisit that claim, using high-resolution simulations capable of tracking the way the small-scale structure reionizes, and using realistic X-ray preheating models.

Throughout this paper we are interested in the observational signatures that cosmic dawn astrophysics (in this case X-ray preheating) leaves on the 3D \lya forest power spectrum via the impact of X-rays on the imprints from \ion{H}{I} reionization that survive in the IGM to low redshifts ($z < 4$). Before covering our implementation of X-ray preheating, we briefly describe the fundamental ideas behind the reionization relics that make it possible to have any X-ray preheating signatures. 

Inhomogeneous reionization seeds temperature fluctuations, fluctuations in the pressure smoothing, and affects the ionizing radiation field \citep[see, e.g.][]{2022MNRAS.tmp.3519P}. The resulting smoking gun is the large-scale enhancement present in the \lya forest power spectra (\citetalias{2019MNRAS.487.1047M}; \citealt{2019MNRAS.486.4075O,2019MNRAS.490.3177W,2022MNRAS.509.6119M})\footnote{\cite{2023MNRAS.521.1489M} presented evidence for a possible measurement at $z \geq 4$.}. However, accurately modeling the imprint of reionization relics on the low-redshift post-reionization IGM ($z \leq 4$) requires great mass resolution because the gas in underdense regions proves crucial. Gas in minivoids gets heated by the UV ionization front to $\sim 2 \times 10^4$ K. In addition, this underdense gas can be shock-heated by shocks that originate in denser regions. As a result, the underdense gas is compressed to the median density and gets heated to even higher temperatures than gas in denser regions \citep{2018MNRAS.474.2173H}. This additional heating, which originates on the way the small-scale structure reionizes and takes cosmological timescales to relax, is the reason for the memory of reionization in the low-redshift ($z \sim 2$) \lya forest. 

To compute the \lya forest 3D flux power spectrum, including the imprints from \ion{H}{I} inhomogeneous reionization, we will use the model described in \S\ref{ssec:3d}, which consists of using the linear matter power spectrum, the biasing model and non-linear correction from \citet{2015JCAP...12..017A}, and a simulation-based hybrid method to obtain the reionization relics. The reason for the hybrid methodology is that the dynamical range is too large for a single simulation because we need not only the statistical power to capture the patchy nature of reionization but also great mass resolution to accurately track the way the small-scale structure reionizes.

\cite{2018MNRAS.474.2173H} established the impact of X-ray preheating on the BAO signal of the \lya forest using a slightly unrealistic X-ray preheating model, where the transition from absorption to emission of the 21~cm brightness temperature occurs at $z < 16$, while the EDGES collaboration could have discovered the absorption peak of the signal centered around $z = 17$ \citep{2018Natur.555...67B}\footnote{The cosmological nature of this result remains controversial (see, e.g., \citealt{2022NatAs...6..607S}).}.  For reference in our analysis, we include the X-ray model introduced in \citet{2018MNRAS.474.2173H}, hereafter referred to as ``Hirata-300 model'', in which the additional injected energy into the IGM due to X-rays is given by
\begin{eqnarray}
    \label{eq:h18}
    \Delta T^{\rm X}_{\rm gas} (z) = 300  \left(\frac{10}{1+z}\right)^{6.5} \ \ \ \ {\rm K} \, .
\end{eqnarray} 

In order to investigate the impact of X-ray preheating in the imprints from patchy reionization, we utilize the  X-ray prescription used in {\tt 21cmFASTv1.3} \citep{2007ApJ...669..663M,2011MNRAS.411..955M}, where the heating rate per particle is computed by summing the contributions from sources located in concentric spherical shells. We refer the interested reader to \cite{2011MNRAS.411..955M} for the explicit prescription. Here we only recap the steps that highlight the free parameters in {\tt 21cmFAST}.

The total X-ray emission rate is a necessary ingredient to compute the X-ray heating per baryon in the pre-reionization era. This rate per redshift interval from luminous sources between $z''$ and $z'' + dz''$ is 
\begin{eqnarray}
    \label{eq:zeta_X}
    \frac{d \dot{N}_{X}}{dz''} = \zeta_{X} f_* \Omega_b \rho_{\rm crit,0} (1 + \delta_{\rm nl}^{\rm R''}) \frac{dV}{dz''} \frac{df_{\rm coll}}{dt}\, ,
\end{eqnarray}
in units of ${\rm s}^{-1}$, where $\zeta_X$ is the number of X-ray photons per solar mass, which manage to escape their parent galaxies, and the remaining terms in the right-hand-side of Eq.~(\ref{eq:zeta_X}) altogether correspond to the total star formation rate inside a spherical shell of inner (outer) radius at the comoving redshift $z''$ ($z'' + dz''$). $f_*$ is the fraction of baryons that becomes stars, $f_{\rm coll}$ is the collapsed fraction, $\Omega_b$ is the baryon density, $\rho_{\rm crit,0}$ is the critical density (today), and ${\rm R''}$ is the comoving distance between $z'$, the redshift of interest, and $z''$. $\delta_{\rm nl}^{\rm R''}$ is the evolved -- with respect to redshift $z''$ -- density smoothed on scales of ${\rm R''}$.  The X-ray efficiency, $\zeta_X$, is one of the free parameters of the X-ray prescription of {\tt 21cmFASTv1.3}, and its role is to control the degree of X-ray heating prior to reionization. Higher values would eventually cause reionization to occur earlier.  

To compute the arrival rate, i.e. the number of X-ray photons per unit time per unit frequency bin with frequency $\nu$ seen at $(\boldsymbol{x}, z')$, from sources within $z''$ and $z'' + dz''$, {\tt 21cmFAST} assumes that the X-ray luminosity of sources follows a power law of the form, $L \propto (E/E_0)^{-\alpha}$, with $\alpha$ being a constant and $E_0$ being the energy threshold for the lowest energy X-ray photons not absorbed by galaxies, and hence escaping into the IGM. The arrival rate can then be written as a function of the total X-ray emission rate (defined in Equation~\ref{eq:zeta_X}) as follows
\begin{eqnarray}
    \label{eq:E0}
    \frac{d\phi_X (\boldsymbol{x}, \nu, z', z'')}{dz''} = \frac{d\dot{N}_{X}}{dz''} \alpha \, \nu^{-1} \left(\frac{\nu}{\nu_0}\right)^{-\alpha - 1} \left(\frac{1+z''}{1+z'}\right)^{-\alpha - 1} e^{-\tau_X} ,
\end{eqnarray}
where $E_0 = h \nu_0$ and the exponential term accounts for the IGM attenuation for X-ray photons. Finally, the code calculates the X-ray heating per baryon by integrating Eq.~(\ref{eq:E0}) over frequency and $z''$ (see the detailed derivation in \S3.1.2 of \citealt{2011MNRAS.411..955M}).

The energy threshold, $E_0$, is the other free parameter in this X-ray prescription that is allowed to vary in this work. A larger value corresponds to inefficient X-ray heating of the IGM due to more X-ray photons being absorbed by the host galaxies, i.e. fewer photons preheat the IGM, and thus there is a slight delay in the reionization process. 

\begin{table}
    \centering
    \begin{tabular}{cc}
    \hline\hline
    Models & Assumptions \\
    \hline
    Fid & No X-ray preheating \\ 
    Fast-Fid & Default {\tt 21cmFAST} X-ray prescription\\ 
    E1 & Same as Fast-Fid but with $E_0 = 100$ eV\\
    E2 & Same as Fast-Fid but with $E_0 = 1000$ eV\\
    E3 & Same as Fast-Fid but with $E_0 = 1500$ eV\\
    $\zeta_X1$ & Same as Fast-Fid but with $\zeta_X = 1\times10^{56}$ M$_\odot^{-1}$\\
    $\zeta_X2$ & Same as Fast-Fid but with $\zeta_X = 4\times10^{56}$ M$_\odot^{-1}$\\
    $\zeta_X3$ & Same as Fast-Fid but with $\zeta_X = 8\times10^{56}$ M$_\odot^{-1}$\\
    Hirata-300 & X-ray prescription from \citet{2018MNRAS.474.2173H}, i.e.\ Eq.~(\ref{eq:h18})\\
    \hline
    \end{tabular}
    \caption{Summary of X-ray preheating models considered in this paper. The default {\tt 21cmFAST} X-ray prescription has  $E_0 = 500$ eV and $\zeta_X = 2\times10^{56}$ M$_\odot^{-1}$.}
    \label{tab:models}
\end{table}

\begin{figure}
    \centering
    \includegraphics[width=\linewidth]{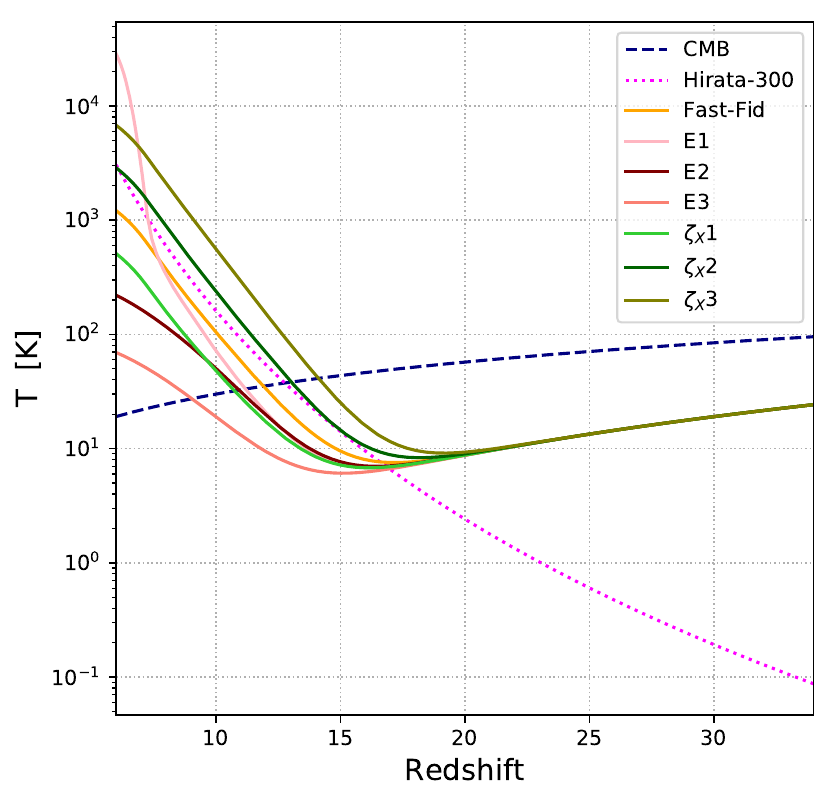}
    \caption{Gas temperature as a function of redshift for all X-ray preheating models considered in this work. We also include the CMB temperature evolution (dashed blue line). Moreover, the Hirata-300 model is also shown (magenta dotted line). The heating transition occurs between $10.7 \lessapprox z \lessapprox 13.0$ for most models.}
    \label{fig:pre-models}
\end{figure}

We use the ``heating models'' of  \citetalias{2020MNRAS.499.1640M} to establish the range of IGM temperatures that we consider throughout this work. In summary, the ``Fast-Fid'' model has $E_0 = 500$ eV and $\zeta_X = 2 \times 10^{56}$ M$_\odot^{-1}$. This should not be confused with the overall Fiducial (``Fid'') model that has no X-ray preheating whatsoever. Furthermore, three variations on the energy threshold are considered $E_0 = \{100, 1000, 1500\}$ eV and three additional variations for the X-ray ionization efficiency $\zeta_X = \{1, 4 , 8\} \times 10^{56}$ M$_\odot^{-1}$. For the nomenclature, we refer to the ``E2'' model as the one with $E_0 = 1000$ eV, the ``$\zeta_X3$'' model has $\zeta_X = 8 \times 10^{56}$ M$_\odot^{-1}$, and similarly for other models. We summarize the nomenclature of models in Table~\ref{tab:models}.

In Figure \ref{fig:pre-models}, we plot the gas temperature evolution with respect to the redshift with different prescriptions of X-ray preheating considered herein, based on the simulations presented in \S3. Note that the $\zeta_X3$ model has the earliest transition from 21~cm absorption to emission ($z \approx 15$). The bulk of models experiences the heating transition from $10.7 \lessapprox z \lessapprox 13.0$, while the latest model to surpass the CMB temperature is the E3 model where it happens at $z = 9$. This feature is expected for the E3 model since most X-ray photons are absorbed by their host galaxies in this photon-starved scenario, therefore leading to a delayed epoch of heating.

From Figure \ref{fig:pre-models}, we observe that, at $z = 10$, the gas temperature ranges from $19$ to $564$ K. We highlight that the high-temperature end is consistent with the findings of \cite{2017MNRAS.468.3785R} and the models with the early heating transition of \cite{2014Natur.506..197F}, while the low end is roughly consistent, albeit slightly earlier for several models, with the current constraints from \cite{2022arXiv221004912T}. Interestingly, the E1 model showcases some slightly deviant behavior with respect to the other models during the epoch of reionization. We attribute this feature to the photon-abundant nature of the X-ray model; however, due to the lower X-ray efficiency compared to that of model $\zeta_X3$ it takes a longer time for those X-ray photons to make their impact. Besides, they get a boost on their escape efforts thanks to the UV ionization going on during reionization.   

The {\sc 21cmFAST} simulation suite of \citetalias{2020MNRAS.499.1640M} computes the gas temperature up to $z \approx 35$. Hence, to ensure that X-rays do not permeate the whole dark age \citep{2012RPPh...75h6901P}, the high-redshift evolution of the gas temperature contribution due to X-rays is assumed to follow a power law \citep{2018MNRAS.474.2173H} as follows. For $1+z>40$, 
\begin{eqnarray}
    \label{eq:knee}
    \Delta T^{\rm X}_{\rm gas}(z) = \Delta T^{\rm X}_{\rm gas} (z = 39) \left(\frac{40}{1 + z} \right)^{3.5}\,.
\end{eqnarray}
Note that Eq.~(\ref{eq:knee}) does not have the same redshift evolution as the Hirata-300 model and, because of the low values of the added gas temperature, the impact on the hydrodynamics outside of the epoch of reionization and epoch of heating will be negligible. For instance, $\Delta T^{\rm X}_{\rm gas}$ is one (two) order of magnitude lower than $T_{\rm CMB}$ at $z = 50$ ($z = 100$). The reason for choosing a different redshift evolution than the Hirata-300 model is because Figure \ref{fig:pre-models} shows that the {\tt 21cmFAST} models do not decay strongly during the cosmic dawn, hence we chose a slightly less sudden drop. Given the large difference between the injected temperature of the gas (due to X-rays) and the CMB temperature at those high redshifts, the choice of \emph{knee} is unlikely to play any role in our analysis --- as long as there is one to limit the role of X-rays in the dark age.

\section{Simulations}
\label{sec:meth}
We follow \citetalias{2019MNRAS.487.1047M}; \citetalias{2020MNRAS.499.1640M}; \citealt{2022arXiv221002385L} on simulating the impact of inhomogeneous reionization in the post-reionization IGM via a hybrid strategy that uses small box high-resolution simulations, which can accurately track down how gas reionizes \emph{below} the Jeans mass \emph{prior} to the passage of an ionization front, and large box semi-numerical simulations, which account for the patchy nature of reionization. We use a modified version of {\tt Gadget-2} \citep{2001NewA....6...79S,2005Natur.435..629S,2018MNRAS.474.2173H} for the small boxes and {\tt 21cmFAST} \citep{2007ApJ...669..663M,2011MNRAS.411..955M} for the large boxes. 

The large box simulation suite was taken from \citetalias{2020MNRAS.499.1640M} and consists of {\tt 21cmFASTv1.3} simulations with a box size of 400 Mpc on each side, with $768^3$ cells for the matter field, and $256^3$ cells for the \ion{H}{I} field. For each heating model considered in this work, we run {\tt 21cmFAST} to generate the IGM temperature, neutral hydrogen field, and density contrast evolution during the epoch of reionization and cosmic dawn. The neutral hydrogen field and density contrast are needed to incorporate the inhomogeneous nature of reionization when calculating the imprint of reionization on the \lya forest (see Equation~\ref{eq:pmpsi} below) while the gas temperature is needed as an input for the {\tt Gadget-2} simulations.

Our {\tt Gadget-2} simulations include adiabatic expansion (including Hubble expansion), shock heating, Compton heating and cooling for neutral gas (accounting for residual ionization). For ionized gas, we also include Compton cooling, \ion{He}{II} cooling, recombination cooling, photoionization heating, and free-free cooling. For details, see \S 4.3 of \cite{2018MNRAS.474.2173H}.

We use two different configurations for the small box simulations --- ``Type I'', a smaller configuration used to swiftly explore the qualitative impact of different X-ray models (see \S\ref{sec:igm} below), and ``Type II'', a larger configuration for comparison purposes using a couple of selected models (see \S\ref{sec:II} below). Both configurations use high-mass resolution simulations which were extensively tested and described in \cite{2018MNRAS.474.2173H}. The Type I simulations have a box size of 425 kpc on each side, $2 \times 128^3$ total number of particles, streaming velocities between baryons and dark matter of 33 km/s (r.m.s. at decoupling), a dark matter particle mass of $1.21 \times 10^3$ M$_\odot$, and a gas-particle mass of $2.27 \times 10^2$ M$_\odot$. 
The Type II simulations have a box size of 1275 kpc on each side, $2 \times 192^3$ total number of particles, streaming velocity of 33 km/s , a dark matter particle mass of $9.72 \times 10^3$ M$_\odot$, and a gas-particle mass of $1.81 \times 10^3$ M$_\odot$. Note that the Type II simulations have been included in this analysis since the Type I boxes are not large enough to reliably sample the linear regime at the redshift of interest for a \lya forest survey. In contrast, the Type II boxes are large enough to form multiple \lya clouds and thus provide enhanced statistical power. Naturally, the ideal scenario would have been to have a single simulation cover both the small-scale structure response to the reionization process and the inhomogeneous nature of reionization. Nevertheless, the dynamical range would be too large to be computationally affordable.  

We ran four different realizations of each Type I and Type II model considered throughout this work. We neglect the variance on the {\tt 21cmFAST} simulations since previous work found that the error budget is dominated by the {\tt Gadget} simulations \citepalias{2019MNRAS.487.1047M}. Although we have computed the sample variance on the mean, for most of our Type I results we do not include this in our plots since the plots are too clustered due to the many models being considered. In contrast, we illustrate the Monte Carlo scatter due to the different realizations of the Fiducial and Fast-Fid model in \S\ref{sec:II}.

For a description of the heating and cooling processes that are implemented in the small box simulations, we refer readers to \S4 of \cite{2018MNRAS.474.2173H}. Here we only recap the inclusion of X-ray preheating. 

In {\tt Gadget-2}, all heating and cooling rates are written as 
\begin{eqnarray}
    \label{eq:rates}
    \frac{\dot{A}_i}{A} = \frac{2 \dot{E}_i}{3 n k_{\rm B} T} \, ,
\end{eqnarray}
where $k_{\rm B}$ is the Boltzmann's constant, $n$ is the total number density of particles, $T$ is the gas temperature, $\dot{E}_i$ is the net volumetric heating rate in erg cm$^{-3}$ s$^{-1}$ from process $i$, $A = p \rho^{-5/3}$ is the particle entropy in the code, and $\dot{A}_i/A$ is the fractional rate of increase of thermal energy from process $i$.

To include X-ray preheating into the {\tt Gadget-2} simulations, it is sufficient to use Eq.~(\ref{eq:rates}) to account for the extra heating $\Delta T$ due to X-rays, i.e. $\dot{A}_X / A \propto \dot{\Delta T_X} / T$. Thus, one needs to replace $\Delta T_X$ using Eq.~(\ref{eq:h18}) in the Hirata-300 model or use results generated with {\tt 21cmFAST} in other models. Note that Eq.~(\ref{eq:knee}) is used here for every {\tt 21cmFAST} model to guarantee that X-rays do not permeate the dark age, i.e. predominant X-ray sources at very high redshifts that could compete with the CMB should not exist. 

\begin{figure*}

     \centering
     \includegraphics[width=\linewidth]{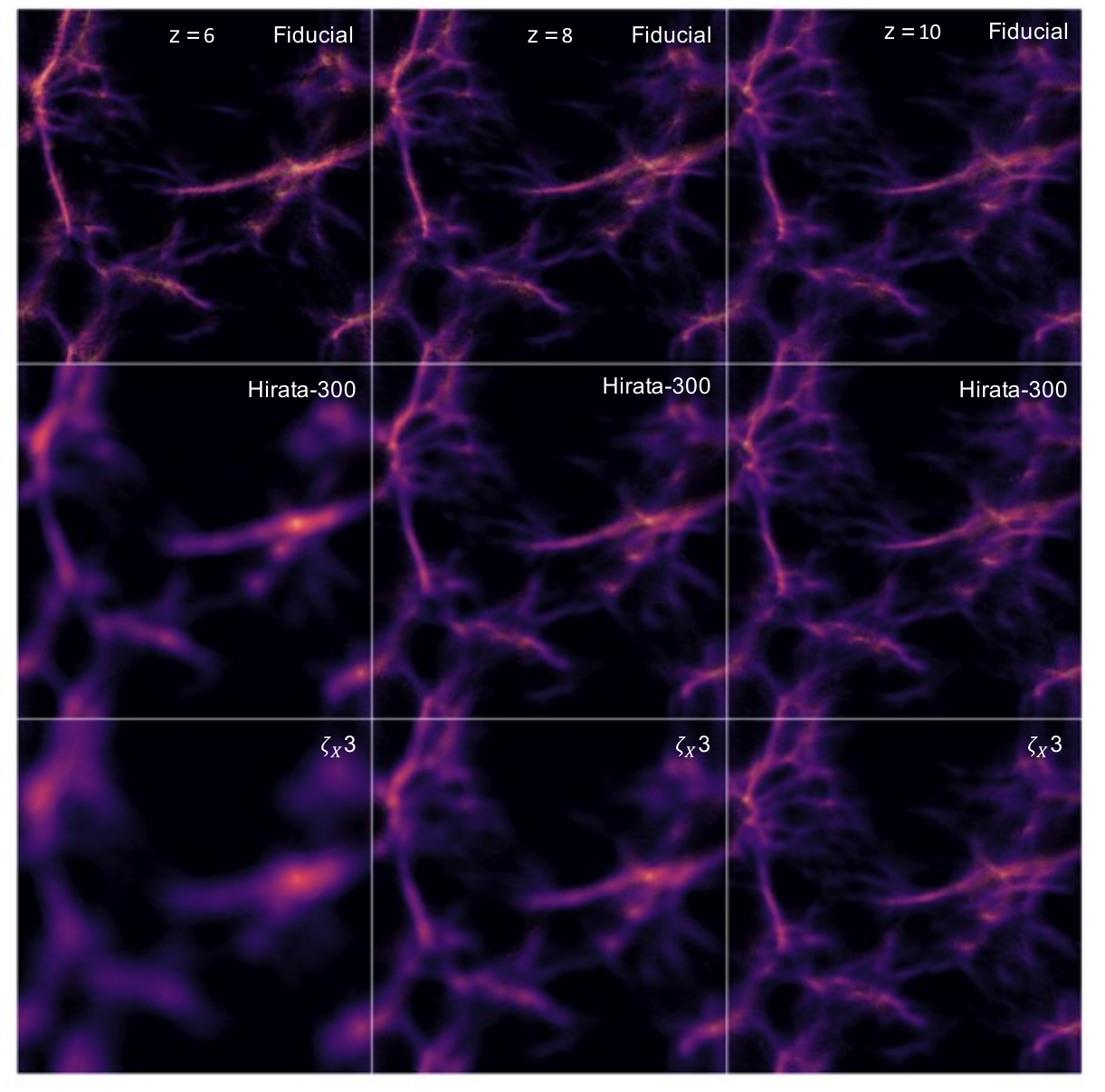} 
     \caption{Log-normal gas density distribution of representative snapshots from the high-resolution Type I simulations, which have a box size of 425 kpc. The left, middle, and right columns correspond to redshifts $z=6$, 8, and 10, respectively. 
     The top, middle, and bottom rows correspond to the Fiducial model (no X-ray preheating of the IGM), 
     the Hirata-300 model (with X-ray preheating prescription in Eq.~\ref{eq:h18}), 
     and the $\zeta_X3$ model (with the X-ray preheating obtained using the {\tt 21cmFASTv1.3} prescription),  respectively.
     } 
    \label{fig:snapshots}
\end{figure*}

In Figure \ref{fig:snapshots}, we plot the gas distribution for our overall Fiducial model, the Hirata-300 model, and the $\zeta_X3$ model at different redshifts, prior to the start of the local reionization process. It is straightforward even by eye to grasp the impact of X-ray preheating --- without X-ray preheating, the Fiducial model exhibits a much richer structure at all the shown redshifts when compared to the latter two X-ray models. Note that the \emph{skeleton} of the denser regions remains although there is significant smearing in both the Hirata-300 and the $\zeta_X3$ model. In contrast, the underdense areas have been severely diminished. Regarding the differences between the two X-ray prescriptions, while they can be found by eye, particularly at the lower redshifts, we can easily quantify the differences in terms of the transparency of the intergalactic medium (see \S\ref{sec:igm} below).

As described in \S4.4 of \cite{2018MNRAS.474.2173H}, to compute the \lya transmission from our small-box snapshots we first select one of the axes as the line-of-sight direction (we later average over the results for all three axes). We assign the \ion{H}{I} abundance proportional to $\Delta^2 \alpha_A(T)$ where $\alpha_A$ is the Case A recombination rate and $\Delta = 1 + \delta_{\rm b}$ is the gas density in units of the mean baryon density in the Universe. The neutral hydrogen particles are then interpolated into a grid based on their redshift-space positions. A Gaussian with thermal width corresponding to the temperature $T$ and the mass of the hydrogen atom is used for the line-of-sight interpolation and we allow the Gaussian to wrap up by including periodic-box images due to the small box size. This procedure results in an opacity cube with arbitrary normalization. We obtain the normalization $\tau_1$ by matching our results to the observed mean flux from \cite{2007MNRAS.382.1657K}.

Thus, the relevant observable from the {\tt Gadget-2} simulations is the $\tau_1$ value as a function of the redshift of observation and local redshift of reionization, and what we aim to identify is how $\tau_1$ changes with the different heating models described in Table \ref{tab:models}.

\section{Results with Type I simulations:  Impact of X-ray preheating on IGM }
\label{sec:igm}

\citetalias{2020MNRAS.499.1640M} demonstrated that while X-ray preheating can affect the initial stages of the reionization process, e.g. by raising the global $\overline{x}_{\rm HII}$ to $\sim 0.1$ before UV heating takes over, this effect has a minimal impact on the imprints from inhomogeneous reionization in the post-reionization IGM, and therefore the resulting memory of reionization in the \lya forest, i.e. the large-scale enhancement in the transmitted flux power spectrum, would be capable of probing the astrophysics of reionization without dealing with the degeneracy introduced by cosmic dawn astrophysics. 

However, X-rays can penetrate deeply into the IGM, which can also effectively wipe out some of the small-scale structures prior to the beginning of the reionization process. As seen in Figure \ref{fig:snapshots}, this impact appears to be even larger in the underdense regions. Underdense regions are crucial to establish the strength of the memory of reionization in the \lya forest because the temperature-density relation becomes bimodal due to the underdense regions suffering shock-heating and compression to mean density in addition to UV heating \citep{2018MNRAS.474.2173H}. The impact of X-ray preheating on the temperature-density relation of the IGM was left unexplored in \citetalias{2020MNRAS.499.1640M}. Armed with the simulations herein, we can now investigate in this paper the long-lasting impact of X-ray preheating in the post-reionization IGM via the impact of inhomogeneous reionization (\citetalias{2019MNRAS.487.1047M}), through its effect on the temperature-density relation of the IGM. See Appendix \ref{app:hemd} for the effect of X-ray preheating in the temperature-density relation. 

The fluctuations in \lya flux transmission, including the effect of reionization, are given by (\citetalias{2019MNRAS.487.1047M})
\begin{eqnarray}
    \label{eq:fluct}
    \delta_{\rm F} = (1 + \beta_{\rm F} \mu^2)\, b_{\rm F}\, \delta_m + b_\Gamma \, \psi \,
\end{eqnarray}
where $b_{\rm F}$ is the flux bias, which is negative since the presence of matter implies absorption, $\beta_{\rm F}$ is the redshift-space distortion parameter, $\mu$ is the cosine of the angle of the $\boldsymbol{k}$ vector with respect to the line of sight, $\delta_m$ is the matter density contrast, $b_\Gamma$ is the radiation bias \citep{2015JCAP...12..017A,2018MNRAS.474.2173H} needed here to convert from optical depth fluctuation to flux. The transparency of the IGM, $\psi$, is a function of both redshift of observation $z_{\rm obs}$ and the redshift of local reionization $z_{\rm re}$, and is defined by the relative strength to a fiducial scenario where reionization happens at $z_{\rm re} = 8$ in the Fiducial model (no X-ray preheating), i.e. 
\begin{eqnarray}
    \label{eq:psi}
    \psi (z_{\rm obs}, z_{\rm re}) = \Delta \ln \tau_1 = \ln \left(\frac{\tau_1(z_{\rm obs}, z_{\rm re})}{\overline{\tau}_1 (z_{\rm obs}, z_{\rm re} = 8)}\right) \, .
\end{eqnarray}
Often in \lya simulations, one varies the normalization --- in this case, $\tau_1$ --- which is equivalent to varying the ionizing background, until one obtains the correct mean transmitted flux in our simulations. Specifically, $\tau_1$ 
is the optical depth that must be assigned to a patch of gas with mean density and temperature $T=10^4$ K in order for the mean transmitted flux to match observations. $\psi$ parametrizes the variations of transparency for an opacity cube that suddenly reionizes at $z_{\rm re}$, and is observed at $z_{\rm obs}$, relative to an opacity cube in the reference simulation with no X-ray preheating and that reionizes at $z_{\rm re} = 8$. Note that $\psi < 0 $ implies that the reference simulation is more transparent or equivalently has a higher $\bar{F}$. 

\begin{figure*}
    \centering
    \includegraphics[width=\linewidth]{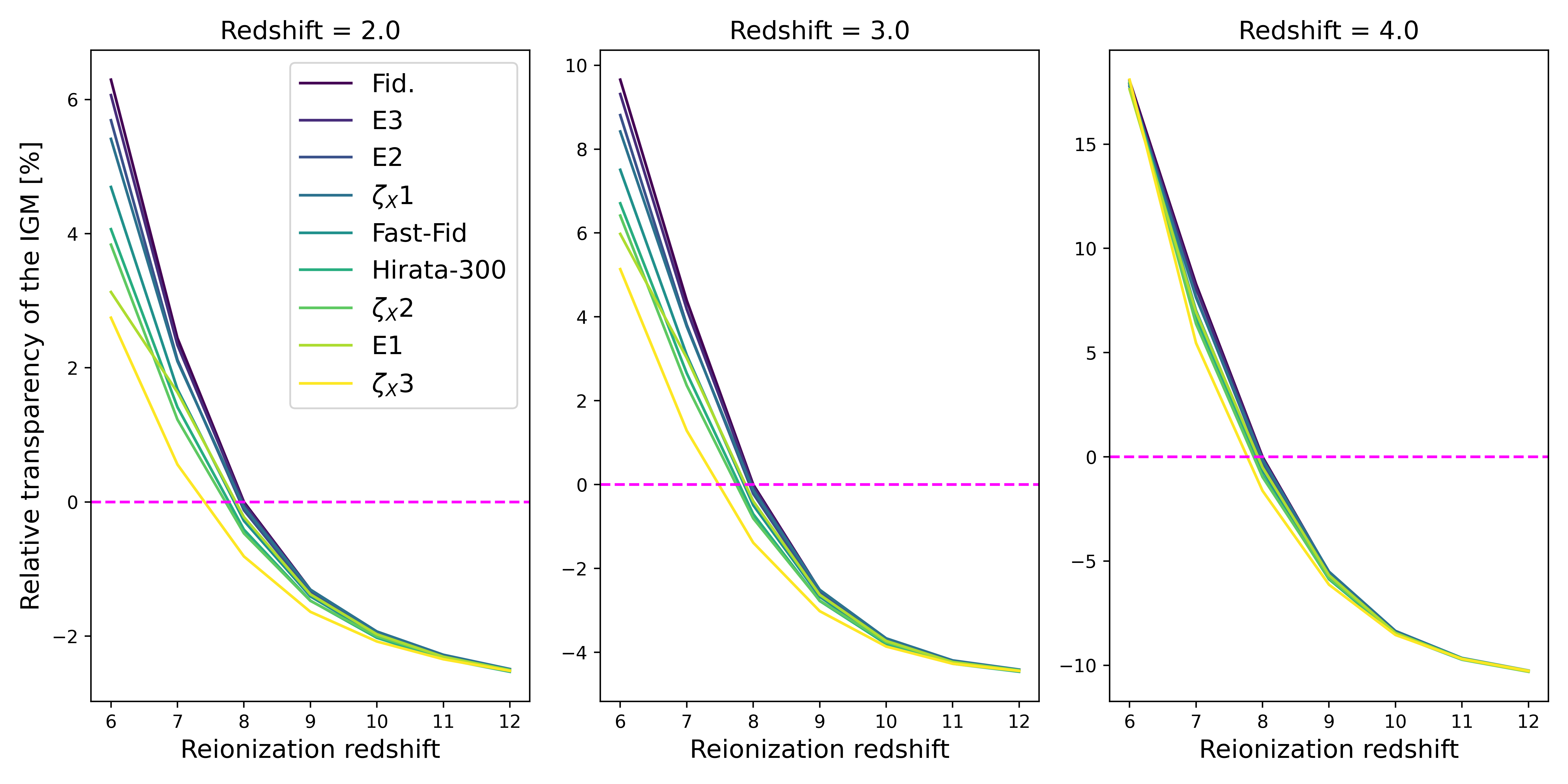}
    \caption{Transparency of the IGM $\psi = \Delta \ln \tau_1$ for our Type I simulations, as a function of local reionization redshift, relative to the scenario where local reionization redshift is $8$ in the Fiducial model. Left, middle, and right columns correspond to different redshifts of observation $z_{\rm obs} = 2.0$, $3.0$, and $4.0$, respectively. For visual reference, we highlight the level of the reference scenario with a dashed magenta line. 
    }
    \label{fig:psi}
\end{figure*}

We plot the relative transparency of the IGM for our Type I high-resolution small-box simulations in Figure \ref{fig:psi}. Irrespective of the redshift of observation, some models like the Fast-Fid and Hirata-300 exhibit a small preference for $\psi^{\rm X-ray} < \psi^{\rm Fid}$ \footnote{This is easier to appreciate in Table \ref{tab:psi}.} for early reionization scenarios ($z_{\rm re} > 8$). Therefore, the consequence of including any X-ray prescription is a less transparent scenario relative to that of the Fiducial model at any $z_{\rm obs}$ for these scenarios. However, as detailed in \S\ref{sec:II}, this trend could be artificial due to the Type I boxes not having a large enough size for enough \lya clouds inside them leading to not-so-accurate sampling. Reassuringly, the different X-ray models, including the simple Hirata-300 model and the Fiducial model, converge for very early reionization ($z = 12$). This can be attributed to two factors. First, the very early start of the reionization process results in less time for X-rays to permeate the IGM. Second, the strong UV radiation field dominates over the X-ray radiation in this case, further limiting the impact of different X-ray prescriptions in this regime. Besides, the gas does not react instantly to X-ray preheating, i.e.\ the \emph{puffing up} of structures takes time, and $z_{\rm re} = 12$ has the shortest response time.   

As seen in Figure \ref{fig:psi}, the same general trend from the early reionization scenarios, $\psi^{\rm X-ray} < \psi^{\rm Fid}$, is present for late reionization scenarios. In this case, the trend is more pronounced and is still generally insensitive to the redshift of observation. The consequence of adding X-rays to our Type I simulations is that the IGM becomes less transparent compared to the Fiducial scenario at a given local redshift of reionization. However, these trends are partially hindered by the fact that the Type I simulations do not accurately sample the forest (see \S\ref{sec:II}). 
In Figure \ref{fig:psi}, the magenta-dashed line indicates where the equality with the reference scenario occurs. The first X-ray model to approach equality is the E3 model at $z_{\rm re} \approx 7.99$, while the last model is the $\zeta_X3$ model that crosses the threshold at $z_{\rm re} \approx 7.5$. These values vary slightly depending on the redshift of observation; however, the order of the models does not change.

For clarity, we tabulate the data used in Figure \ref{fig:psi} in Table \ref{tab:psi}, including redshifts that were omitted in the figure. From Table \ref{tab:psi}, the largest discrepancy between the X-ray prescriptions and the Fiducial model occurs at $z_{\rm re} = 6$ and $z_{\rm obs} = 2.5$, where the Fiducial model predicts that the IGM is 7.840\% more transparent than the reference scenario (i.e.\ Fiducial model with $z_{\rm re} = 8$), compared to 3.489\% more transparent in the $\zeta_X3$ model than the reference scenario, i.e.\ a reduction by more than a factor of two. On the other hand, the smallest deviations tend to occur in early reionization scenarios due to the less prominent role of X-ray preheating in these settings. 

An interesting feature of Figure \ref{fig:psi} is the larger transparency of the fiducial model compared to that of the X-ray models. In principle, one would have expected that the increase in temperature should smooth things out initially since the \lya optical depth decreases with increasing temperature \citep{2001ApJ...549L..11M,2005MNRAS.357.1178B}. We note that the Type-II boxes do recover this expected behavior as seen in \S\ref{sec:II}.

We now move on to quantify the effect of X-ray preheating on the observables that we are considering in this work --- the 3D and 1D flux power spectrum. 

\subsection{3D Flux power spectrum}
\label{ssec:3d}
The \lya flux power spectrum can be computed from Eq.~(\ref{eq:fluct}) by 
\begin{equation}
    \label{eq:P3D}
    P^{\rm 3D}  =   b_{\rm F}^2 (1 + \beta_{\rm F}\, \mu^2)^2\, P^{\rm L}_{\rm m}\, D_{\rm NL} + 2 b_{\rm F} b_\Gamma (1 + \beta_{\rm F} \mu^2) P_{m,\psi} \, ,
\end{equation}
where the first term of the r.h.s.\ is the \emph{conventional} \lya forest power spectrum $P_{\rm F}^{\rm 3D}$, the second of the r.h.s.\ is the contribution due to the memory of reionization $P_{\rm re}^{\rm 3D}$, and we neglect higher order terms of $\psi$. Here,  $P_{m,\psi}$ is the cross-power spectrum of matter and transparency of the IGM (see below for the explicit form), $P^{\rm L}_{\rm m}$ is the linear matter spectrum, which we compute using {\tt CLASS} \citep{2011arXiv1104.2932L,2011JCAP...07..034B}, and the non-linear correction $D_{\rm NL}$ is given by \citep{2015JCAP...12..017A}
\begin{eqnarray}
    \label{eq:Dnl}
    \ln D_{\rm NL} = q_1 \Delta_L^2(k,z) \left[1 - \left(\frac{k}{k_v}\right)^{a_v} \mu^{b_v} \right] - \left(\frac{k}{k_p}\right)^2 \, ,
\end{eqnarray}
which takes care of the isotropic non-linear enhancement (first term on the r.h.s.\ of Eq.~(\ref{eq:Dnl})), Jeans smoothing\footnote{Note that the ``puffing-up'' of structures caused by X-ray preheating is likely to affect the Jeans smoothing of gas. However, this effect would happen at significantly smaller scales than the peak of the reionization relics (\ion{H}{II} bubble scale) and hence we ignore this dependence.}, i.e.\ gas pressure isotropically suppresses the power to below the Jeans scale (third term therein), and the suppression due to peculiar velocities along the line of sight (second term therein). $\Delta_L^2$ is the dimensionless linear matter power spectrum and the different free parameters are taken from Tables 8 and 9 of \cite{2015JCAP...12..017A}, corresponding to simulations with box size $L=60\; h^{-1}{\rm Mpc}$. They also run various simulations to test the convergence of parameters when varying box size and resolution. According to their Tables 2 and 3, when doubling the box size, the changes are within 5 percent for $b_{\rm F}$ and $\beta_{\rm F}$ and within 10 percent for non-linear correction parameters except Jeans scale $k_p$ (about 20 percent). The behaviour is similar when increasing the resolution by a factor of $(5/3)^3$, but due to the limited variation of simulations, their results do not show clear convergence with resolution. However, our analysis of the impact of X-ray preheating on the Ly$\alpha$ power spectrum is unlikely to be significantly affected by the parameter uncertainties.  A rudimentary estimate indicates that the fractional uncertainty is approximately the fractional uncertainty of the parameters, thus about several percent since the X-ray preheating impact mainly arises on large scales. The main source of uncertainty should still be the transparency results from our small-box simulations\footnote{See the discussion in Appendix \ref{app:I}.} and the uncertainty in the timeline of reionization.

The range of redshift in the \cite{2015JCAP...12..017A} tables is from 2.2 to 3.0, so for $z>3$ calculations, we introduce a redshift evolution factor $[(1+z)/(1+z^*)]^{3.55}$ \citep{2013A&A...559A..85P} and use a pivot redshift $z^*=3$. For the conventional Ly$\alpha$ forest power spectrum, we have $P_{\rm F}^{\rm 3D}(z>3)=[(1+z)/(1+z^*)]^{3.55}P_{\rm F}^{\rm 3D}(z^*)$. Analogously, to calculate the memory of reionization term $P_{\rm re}^{\rm 3D}$, the redshift evolution of the flux bias and redshift-space distortion parameter, for $z>3$, is given by 
\begin{equation}
    b_{\rm F}^2(z) (1 + \beta_{\rm F}(z) \mu^2)^2=\left(\frac{1+z}{1+z^*}\right)^{3.55}\left(\frac{P_{\rm m}(z^*)}{P_{\rm m}(z)}\right) \, b_{\rm F}^2(z^*)(1+\beta_{\rm F}(z^*) \mu^2)^2 \, .
\end{equation}
Note that similar bias evolution strategies are often used when dealing with \lya forest observations \citep[see e.g.][]{2020ApJ...901..153D}. 

The second term of Eq.~(\ref{eq:P3D}) corresponds to the memory of reionization, i.e. the imprints due to \ion{H}{I} inhomogeneous reionization in the \lya forest, and the key physical quantity is the cross-power spectrum of matter and transparency of the IGM, which parametrizes the role of inhomogeneous reionization -- primarily, the \emph{void response} to the reionization process. It can be written as (see \citetalias{2019MNRAS.487.1047M} for the derivation)
\begin{eqnarray}
    \label{eq:pmpsi}
    P_{m,\psi}(k,z_{\rm obs}) = - \int {\rm d}z \frac{\partial \psi}{\partial z}(z,z_{\rm obs}) P_{m,x_{\rm HI}}(k,z)\frac{D_g(z_{\rm obs})}{D_g(z)}\, ,
\end{eqnarray}
where the integral covers the epoch of reionization and cosmic dawn, $D_g$ is the growth rate, $P_{m,x_{\rm HI}}$ is the cross-power spectrum between matter and neutral fraction field, which reflects the ionized bubble spatial structure and is needed to account for the patchy nature of reionization. Here, $\partial \psi/ \partial z_{\rm re}$ quantifies the response of the IGM to the reionization process, i.e.\ how the transparency of the IGM changes as the local redshift of reionization changes. Note that each preheating model gives us a unique $P_{m,\psi}(k, z_{\rm obs})$ that accounts for the evolution of both $P_{m,x_{\rm HI}}(k, z_{\rm re})$ and $\partial \psi/\partial z_{\rm re} (z_{\rm re}, z_{\rm obs})$ during the epoch of reionization.  

One final caveat before computing the long-lasting effect of X-rays in the IGM is that we include a biasing procedure when computing $P_{m,\psi}$ at the largest scales. The need for this procedure is clear because our {\tt 21cmFAST} simulations only have a box size of 400 Mpc, and so there are just so few modes that can fit in the first few bins when the separation between the bins is $\Delta k = 2\pi / 400 $ Mpc$^{-1}$. Therefore we implement a cutoff at $k_{\rm cut} = 0.044$ Mpc$^{-1}$. At $k < k_{\rm cut}$, we model $P_{m,\psi}$ using a linear biasing model,
\begin{eqnarray}
    \label{eq:bias}
    P_{m,\psi}(k,z) = \frac{P_{m,\psi}(k_{\rm cut},z)}{P_{\rm m}^{\rm L}(k_{\rm cut},z)} P_{\rm m}^{\rm L}(k,z) \, ,
\end{eqnarray}
which should be accurate at scales larger than the ionization bubble scale. 

\begin{figure*}
    \includegraphics[width=\linewidth]{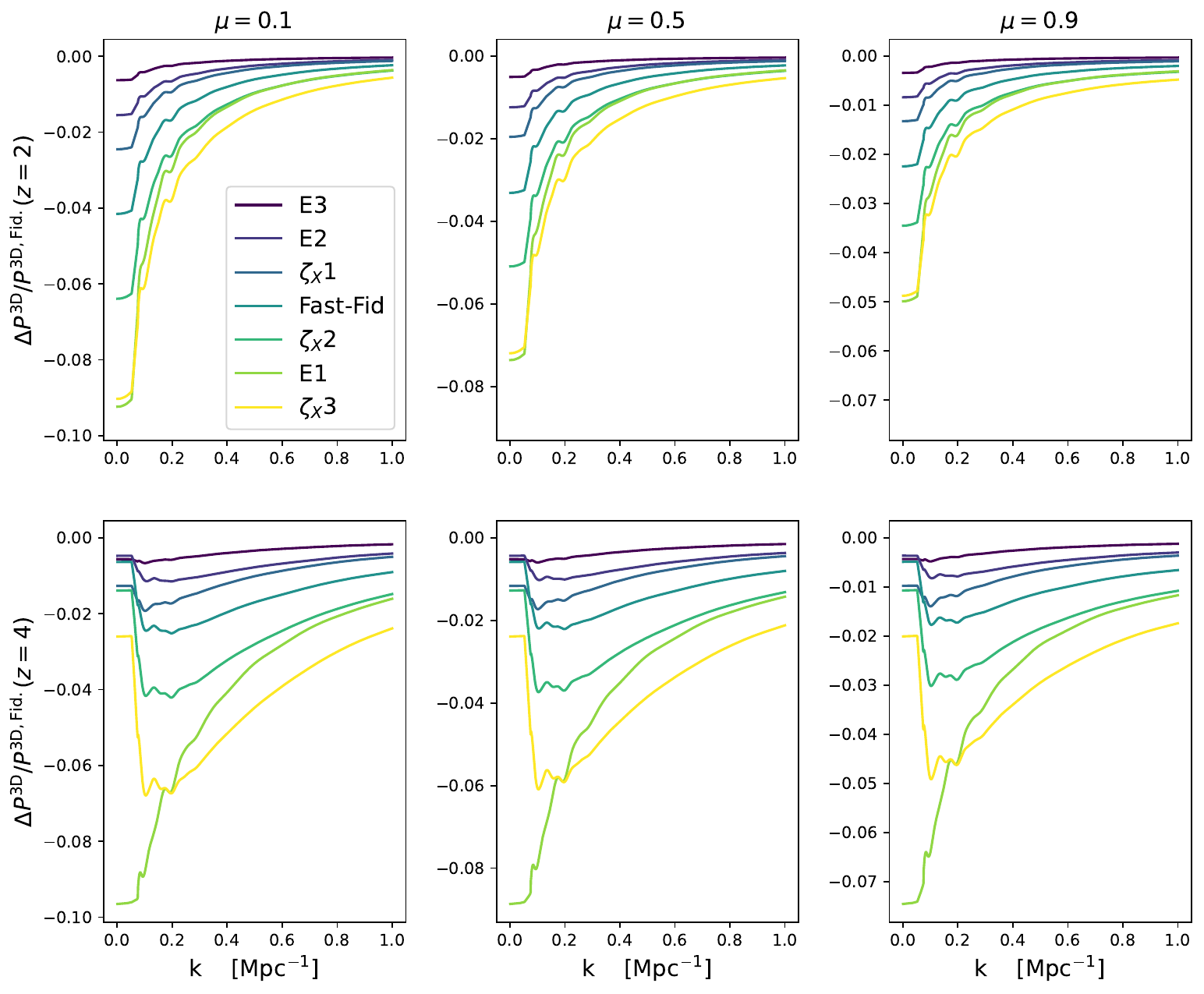}
    \caption{The impact of X-ray preheating in the 3D \lya power spectrum --- the fractional difference of the \lya power spectrum in different X-ray preheating models with respect to that in the Fiducial model. Shown are the results at three different $\mu$ (the cosine of the angle with respect to the line of sight), at low redshift $z_{\rm obs} = 2$ (top) and at high redshift $z_{\rm obs} = 4$ (bottom), respectively. 
    The largest deviation occurs at high redshifts and in the direction nearly perpendicular to the line of sight (i.e.\ small $\mu$). }
    \label{fig:P3D}
\end{figure*}

We chose to quantify the impact of X-ray preheating in the 3D \lya power spectrum in terms of the fractional difference of model ``x'' with respect to the Fiducial model, 
\begin{eqnarray}
    \label{eq:Delta_3D}
    \frac{\Delta P^{\rm 3D,x}}{P^{\rm 3D, Fid}} = \frac{P^{\rm 3D, x}_{\rm re} - P^{\rm 3D, Fid}_{\rm re}}{P_{\rm F}^{\rm 3D, Fid} + P_{\rm re}^{\rm 3D, Fid}} \, .
\end{eqnarray}
The change of pressure smoothing with respect to X-ray preheating is negligible for non-small scales. Consequently, only the reionization term appears in the numerator of the fractional difference. In Figure~\ref{fig:P3D}, we plot the fractional difference using Eq.~(\ref{eq:Delta_3D}) for the different X-ray models at different angles with respect to the line of sight and at two different redshifts. (Note that we drop the Hirata-300 model from the analysis at this point since there is no simple way of implementing a {\tt 21cmFAST} model that has an analogous evolution of the gas temperature.) 

At low redshifts, Figure \ref{fig:P3D} indicates the maximum deviation from the Fiducial model occurs for the E1 model with a roughly nine percent decrease in the signal, although there is intense competition with the $\zeta_X3$ model. Meanwhile, the minimum variation is encountered in the E3 model at large $\mu$ where the impact is quantified at less than a percent. The effect of X-ray preheating is also more substantial in directions perpendicular to the line of sight, which remarkably aligns with the ($\mu$) sweet spot to recover the memory of reionization in the \lya forest because the reionization term -- second term in Eq.~(\ref{eq:P3D}) -- has only one Kaiser term compared to the conventional \lya term that has two Kaiser factors. Overall, the hierarchical structure between the X-ray preheating models can also be appreciated in the impact of the 3D flux power spectrum, with the E3 model being the one with the lowest variation and the $\zeta_X3$ and E1 models competing for the largest one. Note that this trend is consistent with that observed in the transparency of the IGM at $z_{\rm obs} = 2$ (see Figure \ref{fig:psi}) and consistent with the gas temperature (Figure \ref{fig:pre-models}) once one accounts for the fact that the extra heating in the photon-abundant model (E1) has a lower X-ray efficiency compared to that of the $\zeta_X3$ model. Hence the constant early rise of gas temperature is slightly less important than the sudden late rise since there is not enough time to completely influence the transparency results, especially at higher redshifts. Interestingly, the situation ``normalizes'' when observation is at a low enough redshift. Naturally, the scenario is likely to differ if one observes close to the tail end of reionization because the impact of the large amplitude of the gas temperature in the late-rise scenario might become dominant, which can already be seen in the fractional difference at high redshifts of Figure \ref{fig:P3D}. 

Analogously, at high redshift, we see the same trend with respect to $\mu$. Now the most significant difference clearly happens for the E1 model and corresponds to a decrease of roughly 10 percent. However, there is fierce competition and the $\zeta_X3$ model has the largest deviations at some small scales. This sudden takeover at large scales is likely due to the large impact of the last snapshot (for the local reionization scenarios, i.e. $z_{\rm re} = 6$) in the integrand of Eq.(\ref{eq:pmpsi}) -- see also Figure 1 of \citetalias{2020MNRAS.499.1640M}. The smallest change still corresponds to the E3 model in directions close to the line of sight, even though the E2 model can take the last spot at the smallest wavenumber, given the error bars this last feature is likely not statistically significant.      

Although the amplitude of the fractional difference seen in Figure \ref{fig:P3D} for our Type I boxes is somewhat expected because the Fiducial scenario 
disregards any X-ray preheating physics, the spreads between the X-ray preheating prescriptions are surprising. For instance, at $k=0.1$ Mpc$^{-1}$ and $\mu=0.1$ the models span the range of $\sim 0.05 \,P^{\rm 3D, Fid}$ at $z_{\rm obs} = 2$ and $\sim 0.08 \,P^{\rm 3D, Fid}$ at $z_{\rm obs} = 4$. However, note that the significance of the spread is slightly diminished if one were to include X-ray preheating physics in the Fiducial model. In this case, the Fast-Fid model would be the reference point and therefore the difference between the X-ray models will be slightly lesser but still quite significant. For now, the choice of this new Fiducial scenario is a good compromise between modeling accuracy and computational resources for future studies aiming at quantifying the response of the IGM to the reionization process. The sensitivity to X-ray preheating opens a novel window into the astrophysics that governs the cosmic dawn \emph{billions} of years after it happened.

\subsection{1D Flux power spectrum}
\label{ssec:1d}
As the complement to the 3D flux power, the \lya forest 1D power spectrum requires the cross-correlation between pixels along the same line of sight. Mathematically, the expression is given by integrating the 3D counterpart over the perpendicular direction to the line of sight (\citealt{2013A&A...559A..85P}; see also Eq.~10 of \citetalias{2019MNRAS.487.1047M})
\begin{eqnarray}
    \label{eq:p1d}
    P^{\rm 1D}  &=&  \int_0^{\infty} \frac{dk_\perp}{2\pi} P^{\rm 3D} \, ,
\end{eqnarray} 
and can be written as the sum of the \emph{conventional} \lya forest power spectrum $P_{\rm F}^{\rm 1D}$ and the contribution due to the impact of inhomogeneous reionization $P_{\rm re}^{\rm 1D}$. Here, we implement a cutoff at $k_{\rm max} = 7 $ Mpc$^{-1}$. The arbitrary choice of this cutoff is not expected to change our results since the 
\ion{H}{I} reionization relics are coupled to the ionization bubble scales and quickly decline as the wavenumber increases, and so it is close to zero at $k = 1 $ Mpc$^{-1}$.

\begin{figure*}
    \includegraphics[width=\linewidth]{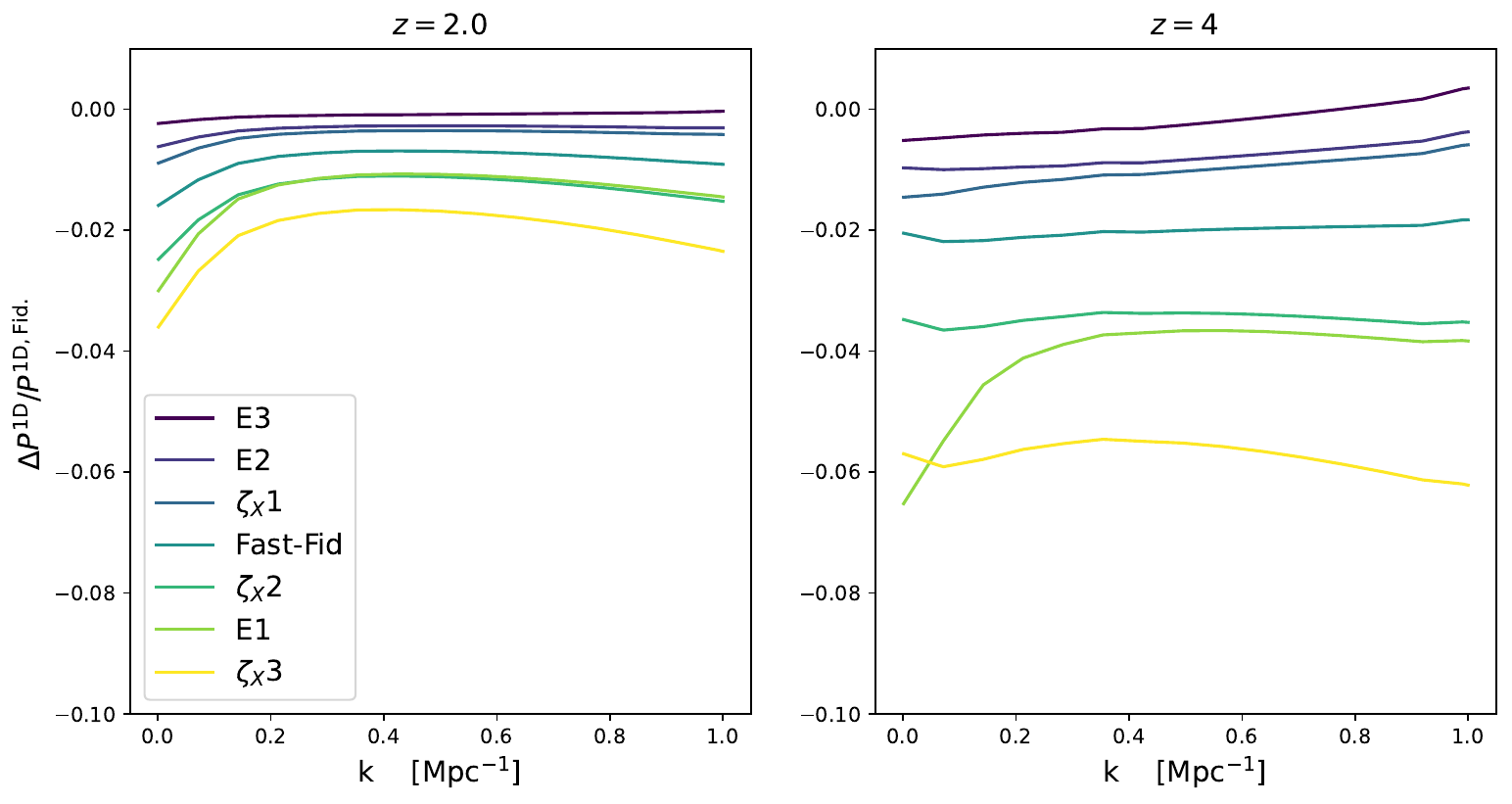}
    \caption{The impact of X-ray preheating in the 1D \lya power spectrum --- the fractional difference of the \lya power spectrum in different X-ray preheating models with respect to that in the Fiducial model. 
    Shown are the results at low redshift $z_{\rm obs} = 2$ (left) and at high redshift $z_{\rm obs} = 4$ (right), respectively. } 
    \label{fig:P1D}
\end{figure*}

Analogous to Eq.~(\ref{eq:Delta_3D}), the 1D fractional difference is
\begin{eqnarray}
    \label{eq:Delta_1D}
    \frac{\Delta P^{\rm 1D,x}}{P^{\rm 1D, Fid}} = \frac{P^{\rm 1D, x}_{\rm re} - P^{\rm 1D, Fid}_{\rm re}}{P_{\rm F}^{\rm 1D, Fid} + P_{\rm re}^{\rm 1D, Fid}} \, .
\end{eqnarray}
We assume that the impact of X-ray preheating on the pressure smoothing is small at the redshift range and scales of interest\footnote{We warn against the use of this approximation at higher redshifts ($z \sim 5$), particularly for $P_F^{\rm 1D}$ analyses because of mode-mixing and the larger role of denser gas in the response of the IGM to the reionization process. Further investigation is required in this domain.}. Hence, the fractional difference for the 1D power spectrum has the same form as the 3D counterpart. In Figure \ref{fig:P1D}, we plot the fractional difference in the 1D flux power spectrum for the different X-ray preheating models considered by our Type I simulations. 

At low redshifts, the 1D fractional difference recovers the hierarchical structure observed in Figure \ref{fig:P3D} with the largest deviation ($< 4\%$ in absolute value) caused by the $\zeta_X3$ model while the smallest effect ($< 1\%$) is present in the E3 model. At high redshifts, we observe a similar order of the models -- albeit with fierce competition between E1 and $\zeta_x3$ models at large scales -- and a larger spread in the values of the fractional differences that is, crucially, comparable to the strength observed in the 3D findings. In particular, we obtain a maximum deviation larger than 6 percent for both the E1 and $\zeta_X3$ models. From an optimistic point of view, this is a promising avenue for learning about the astrophysics that governs the cosmic dawn since medium and high-resolution P1D measurements and spectra already exist \citep{2019JCAP...07..017C,2019MNRAS.482.3458M,2021AJ....161...45O}. On the other hand, this implies that the \lya forest can produce biased results in the absence of proper countermeasures against this broadband systematic. However, as shown in \cite{2023MNRAS.520.4853M}, this effect will not impact the accuracy of the location of the  BAO peak at any redshift but can affect measurements dependent on the shape quite drastically, which could, for instance, jeopardize recent efforts to weaponize the AP effect \citep{1979Natur.281..358A} to extract additional cosmological information from \lya two-point statistics \citep{2022arXiv220912931C,2021MNRAS.506.5439C}. 

Furthermore, the 1D fractional differences have a distinct morphology with respect to the 3D analogs. This trend is easily understood via Eq.~(\ref{eq:p1d}) ---- the integration over perpendicular directions to the line of sight leads to mode-mixing; hence small wavenumbers, which suffer more greatly from the impact of reionization and therefore of X-ray preheating, can be mapped into large wavenumbers. This shape is also encouraging, particularly for future efforts aimed at constraining X-ray preheating astrophysics (e.g.\ properties of the X-ray sources in the high-$z$ IGM) through the impact of inhomogeneous reionization in the 1D flux power spectrum. This shape is also the reason why the $\zeta_X3$ model has an easier time at both redshifts and most scales compared to its rival the E1 model which tends to rise at large scales in $P^{\rm 3D}$.      

Given that $P^{\rm 1D}$ data already exists, future work will assess the potential implications of considering X-ray preheating as part of the fiducial model for inference of cosmological parameters. For now, we highlight that the values seen in Figure \ref{fig:P1D} are larger than the statistical errors of recent \lya forest surveys \citep{2019JCAP...07..017C}. In particular, the statistical error at $k=0.1$ Mpc$^{-1}$ is 0.0033 Mpc at $z=2.2$ and 0.0731 Mpc at $z=4$. In comparison, the effect of X-ray preheating in the IGM (i.e.\ no X-ray preheating vs the $\zeta_X3$ X-ray prescription) at the same wavenumber is 0.0055 Mpc at $z=2.2$ and 0.0654 Mpc at $z = 4$.

Finally, we recognize that the results obtained in \S\ref{ssec:3d} and in \S\ref{ssec:1d}, which use our Type I simulations, suffer from poor sampling given the small box size. Nonetheless, they are sufficient to inform us of the general trends between the different X-ray prescriptions, particularly to highlight the ability to discern between different cosmic dawn astrophysics. To accurately capture the long-lasting impact of X-ray preheating in the flux power spectra, we include the analysis of the larger, Type II boxes in \S\ref{sec:II}, where we only consider the Fast-Fid X-ray prescription and compare it with the Fiducial model due to limited availability in computational resources.

\section{Results with Type II simulations: accurate calculation for the fast-fid model}
\label{sec:II}

In this section, we showcase the impact of X-ray preheating in our Type II boxes, which have a box size of 1275 kpc (a factor of 3 larger than the Type I boxes). These Type II boxes are large enough to form multiple \lya clouds and thus enhance our statistical power. Thus, the Type II simulations more accurately sample the linear regime for the redshifts of interest for \lya forest surveys. Motivated by our findings in \S\ref{sec:igm}, i.e.\ a significant deviation from the fiducial model but a harder-to-resolve discrepancy between X-ray prescriptions, here we focus only on the Fast-Fid X-ray preheating model, which we recommend future work should do when accounting for X-ray preheating signatures present in the post-reionization IGM, namely to consider the fiducial X-ray preheating prescription of {\tt 21cmFAST}. This focus herein is also due to the limitation in computational resources. For reference, we compare with a Type II version of our fiducial model, i.e.\ no X-ray preheating whatsoever. 

For clarity, the only piece of our hybrid strategy being changed in this section is the substitution of the Type I $\psi$ results for the Type II transparencies. Thus, there is no modification for the ``heating models'' from \citetalias{2020MNRAS.499.1640M} that form our large box simulation suite. 

\subsection{Impact on IGM evolution}

\begin{figure*}
    \includegraphics[width=\linewidth]{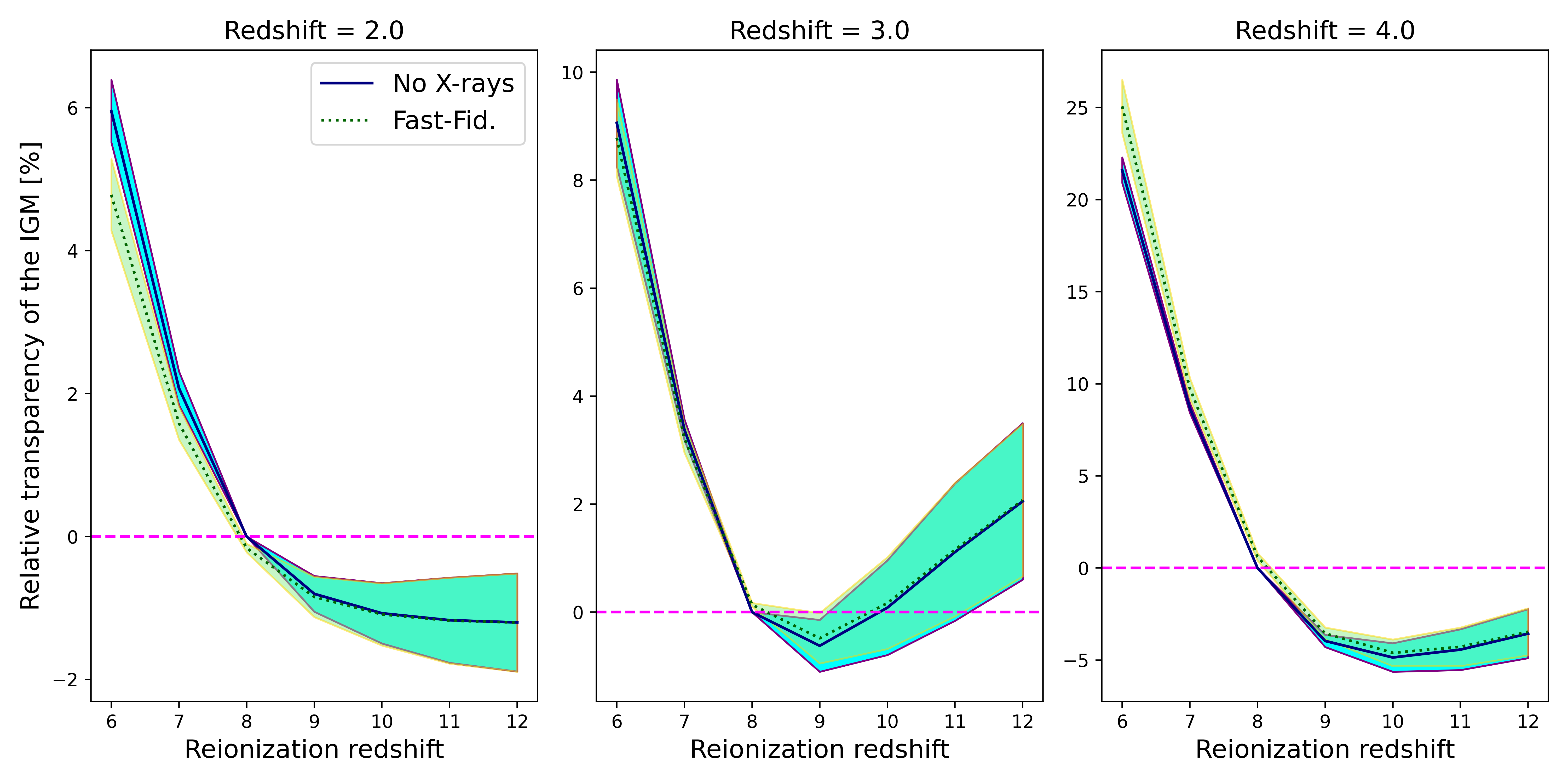}
    \caption{Same as Figure~\ref{fig:psi} but for the Type II boxes and including the Monte Carlo scatter due to different realizations. We only consider the overall fiducial model with no X-ray preheating (blue solid line) and the standard {\tt 21cmFAST} X-ray prescription, ``Fast-Fid'' model (green dotted line). Interestingly, in contrast to Figure \ref{fig:psi}, there is now a turnover with the X-ray model leading to larger transparency initially at high redshifts.}
    \label{fig:psi-II}
\end{figure*}

We tabulate the relative transparency as a function of the redshift of observation and of the redshift of local reionization, which are the data used in Figure~\ref{fig:psi-II}, in Table \ref{tab:psi-II}. Note that the reference scenario is now a Fiducial model Type II box where the local reionization occurs at $z_{\rm re} = 8$. Furthermore, we showcase the transparency results for the Type II boxes in Figure \ref{fig:psi-II}. Figure~\ref{fig:psi-II} shows that the Fast-Fid model, which uses the default prescription for X-ray preheating in {\tt 21cmFAST} \citep{2011MNRAS.411..955M}, is more transparent than the Fiducial model at high redshift of observation and for both late and early reionization scenarios; nevertheless, for early reionization scenarios, the Fiducial and the Fast-Fid model are both less transparent than the reference, which is the same trend observed for the Type I boxes in Figure \ref{fig:psi}.

In contrast to the Type I results, Figure \ref{fig:psi-II} showcases a transition from more transparent to less transparent for the Fast-Fid model as time evolves, at least within the error bars. Hence, we see a recovery of the trend found in Figure \ref{fig:psi} at low redshift of observation and for late reionization scenarios, albeit with slightly lower amplitude for the Fiducial model. For early local reionization, the tendency is obscured due to the Monte Carlo scatter between the different realizations.

Interestingly, the Type II boxes reach considerably larger values than the Type I counterparts at the high redshift of observation. Besides, $\partial \psi / \partial z_{\rm re}$ achieves considerably larger values (in absolute magnitude) than for the Type I boxes at late $z_{\rm re}$. This feature is of significant importance because of the redshift evolution nature of the reionization relics in the forest, where the highest redshifts lead to a stronger signal since the HEMD gas has not had enough time to relax the additional injected energy during the reionization process \citep{2018MNRAS.474.2173H}. Furthermore, the Type II boxes have different behavior at the high end of the local reionization redshift tail. We attribute these features to the increased statistics and more reliable sampling of the linear regime that the Type II boxes achieve compared to that of the Type I analogs. In particular, better sampling allows the Type II boxes to more accurately capture the impact of the initial larger (high $z_{\rm obs}$) gas temperature and, consequently, the wiping-out of small-scale structures. In contrast, at low $z_{\rm obs}$, the initial higher temperature -- and changed densities -- has an easier time relaxing to the temperature-density relation because there should be a lesser HEMD phase.

The maximum deviation, quantified as the absolute difference between the two models, present in our relative transparency results is 3.465\% and occurs at $z_{\rm re} = 6$ and $z_{\rm obs} = 4$, which is smaller than the one predicted for the Type I models (4.518\%). However, this was achieved by the $\zeta_X3$ model at $z_{\rm re} = 6$ and $z_{\rm obs} = 3$, and thus provides an unfair comparison. On the other hand, the minimum deviations tend to occur at early reionization scenarios (and at low redshifts of observation), which is also the trend seen in Table \ref{tab:psi}.

\subsection{Impact on flux power spectra}
Because of the enhancement at high $z_{\rm obs}$ in Figure \ref{fig:psi-II}, we now expect the impact on the \lya forest power spectrum lead to a rise on the X-ray heating signatures at the high redshift of observation, and eventually result in a decrease on the strength of the effect at low redshifts, which is a similar see-saw pattern to the one that is expected in the 21~cm power spectrum \citep{2007MNRAS.376.1680P}. The cause of these see-saw mechanisms is essentially the same: X-ray photons will first increase the power and raise the temperature; however, this would eventually lead to a decrease in power since it comes at the cost of erasing fluctuation in small scales.

\begin{figure*}
    \includegraphics[width=\linewidth]{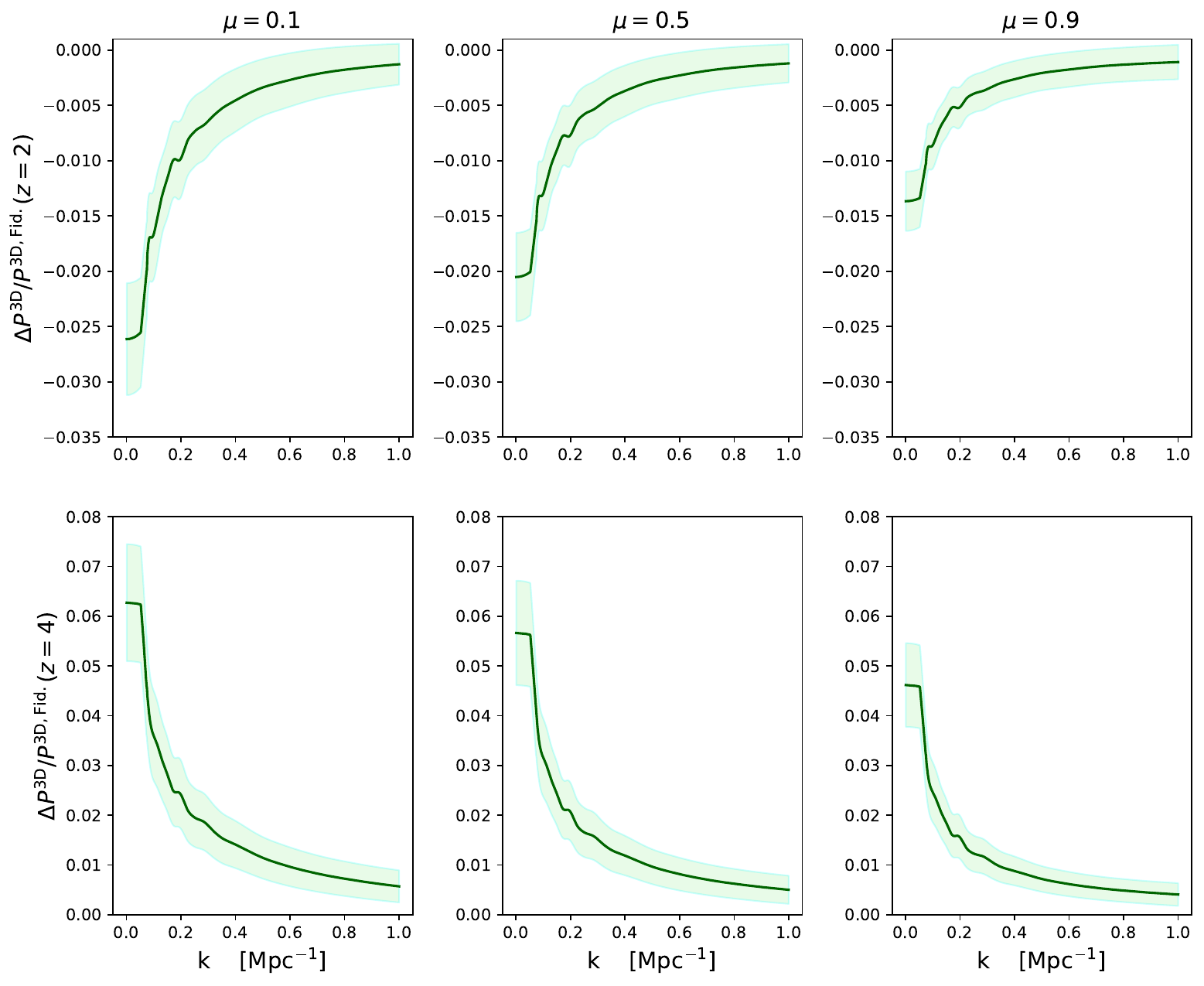}
    \caption{Same as Figure~\ref{fig:P3D} but for the Type II boxes with only the standard {\tt 21cmFAST} X-ray prescription, ``Fast-Fid'' model and including the Monte Carlo scatter due to different realizations. The green line represents the mean result from the different realizations.} 
    \label{fig:P3D-II}
\end{figure*}

We plot the fractional difference of the 3D flux power spectrum for the fiducial {\tt 21cmFAST} model (``Fast-fid'' model) with respect to the model with no X-ray preheating in Figure \ref{fig:P3D-II}. Figure \ref{fig:P3D-II} includes the standard deviation from the mean due to the use of four different realizations of the small-scale high-mass resolution simulations. 

At low redshift of observation, Figure \ref{fig:P3D-II} shows the trend of increasing the impact (in absolute value) at small values of $\mu$, which is the same trend observed in Type I boxes (see Figure \ref{fig:P3D}). Nonetheless, Type II boxes have a slightly diminished impact, with a maximum deviation -- at large scales and along the perpendicular direction -- around three percent. Do note that a potential Type II $\zeta_X3$ model (or even an E1 model) would give rise to an enhancement in the relative strength of the signal. However, the purpose of this section is to establish what can be expected for a fiducial model of the long-lasting impact of X-ray preheating in the \lya forest. Thus, the priority is not on the intriguing ability of this effect to constrain the astrophysics that governs cosmic dawn.  

On the other hand, at high redshift, Figure \ref{fig:P3D-II} shows a strengthening of the memory of reionization -- the imprints from \ion{H}{I} reionization in the \lya forest -- when accounting for the effect of X-ray preheating in the IGM. The effect is reinforced at large scales and for small values of $\mu$, which corresponds exactly to the best window of opportunity to directly capture the epoch of reionization astrophysics from the \lya forest \citep{2021MNRAS.508.1262M}. In particular, we obtain an enhancement of more than six percent in this window. This feature is unique to Type II boxes since it originates on the larger transparency of the X-ray preheating model with respect to the reference scenario at high $z_{\rm obs}$ and for late local reionization.

\begin{figure*}
    \includegraphics[width=\linewidth]{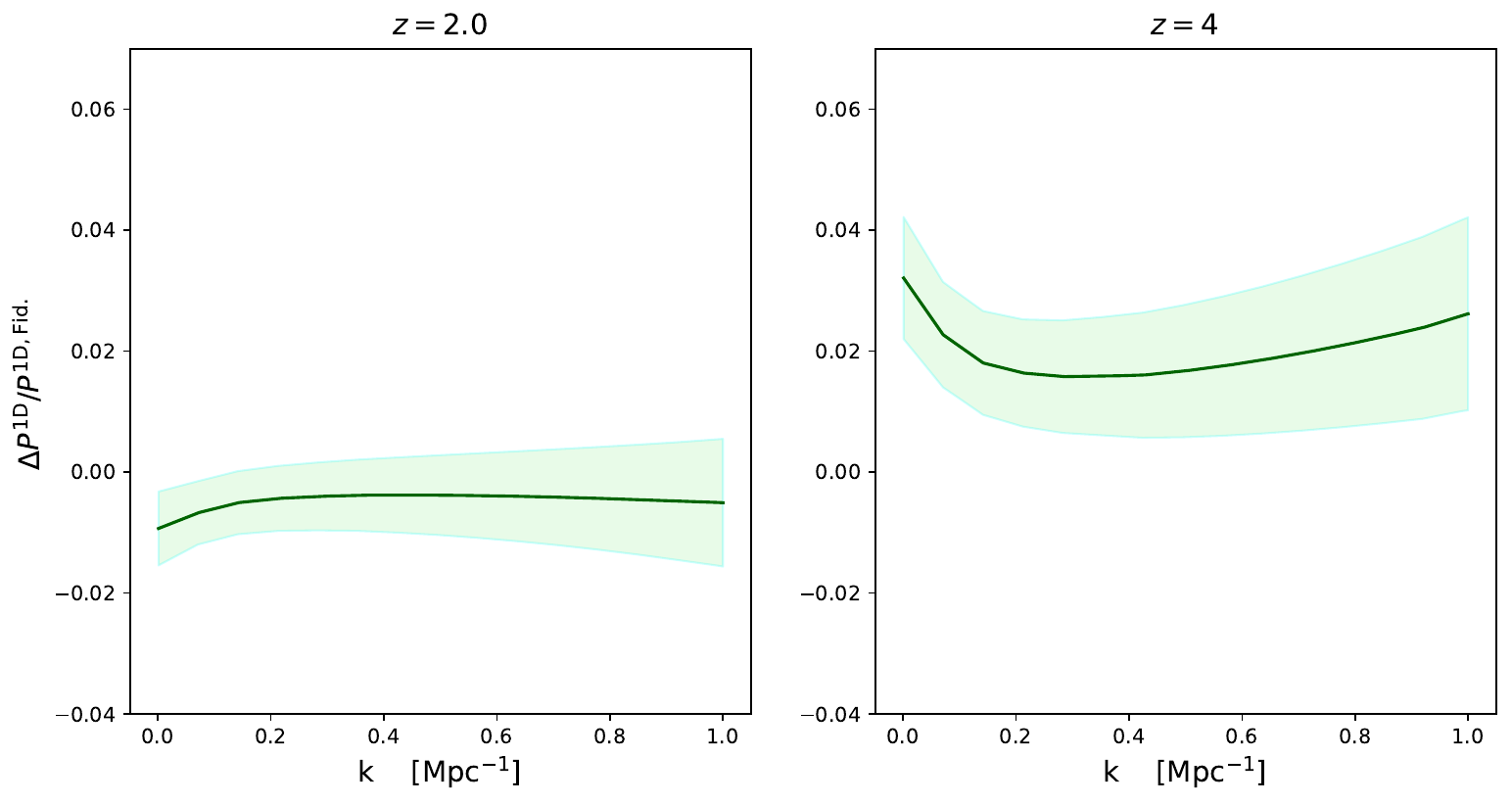}
    \caption{Same as Figure~\ref{fig:P1D} but for the Type II boxes  with only the standard {\tt 21cmFAST} X-ray prescription, ``Fast-Fid'' model and including the Monte Carlo scatter due to different realizations. The green curve shows the mean result.} 
    \label{fig:P1D-II}
\end{figure*}

Likewise, Figure \ref{fig:P1D-II} illustrates the effect of X-ray preheating in the 1D \lya flux power spectrum as a function of wavenumber and both for the low and high redshift of observation. Similar to Figure \ref{fig:P3D-II} the Monte Carlo scatter from different realizations of the high-resolution {\tt Gadget2} simulations has been included. The mean result at low redshift is consistent with the values obtained for Type I boxes (see Figure~\ref{fig:P1D}), although it is also diminished, analogously to the 3D effect. We see a decrease in the signal, with respect to a scenario with no X-ray preheating, peaking at less than two percent.

In addition, at high redshift, we observe a moderate increase in the signal at large scales of less than four percent while a more modest enhancement is also perceptible at smaller scales (slightly larger than two percent). Again, this morphology occurs because of the integration with respect to the directions perpendicular to the line of sight, which can map small $k_\perp$ modes into large $k$ and vice versa. In comparison with Figure \ref{fig:P1D}, the relative impacts seen in Figure \ref{fig:P1D-II} may be lesser but still significant, particularly due to the enhancement of the memory of reionization in the regime where it is the strongest and due to its relative strength when compared with current statistical error \citep{2019JCAP...07..017C} and compared to that of near-term measurements  \citep[e.g.,][]{2023arXiv230606311R,2023arXiv230606316K}.

\section{Summary}
\label{sec:conc}
The high-redshift range of the \lya forest is often seen as an attractive probe of the nature of dark matter, particularly to constrain warm dark matter models \citep[see e.g.][]{2020JCAP...04..038P,2022MNRAS.tmp.3519P}. However, our findings suggest yet another hurdle on the way for the quest for the nature of dark matter. Namely, the thermal state of the high redshift IGM not only depends on the relics from cosmic reionization, but through them, it also depends on the mechanism that set the initial conditions for cosmic reionization, i.e.\ X-ray preheating. 

The precise mechanism through which the early intergalactic medium is preheated prior to cosmic reionization is highly uncertain, which arises, due to a lack of direct observational constraints, from the extrapolation of our understanding of the low-redshift Universe, particularly the X-ray luminosity functions \citep{2022ApJ...930..135L}, and the uncertainty regarding the abundance, clustering, and nature of the X-ray sources \citep{2021ApJ...912..143M}. Consequently, new avenues of probing the cosmic dawn are extremely valuable opportunities for discerning the astrophysics that governs this process. In this work, we establish, for the first time, the long-lasting impact of X-ray preheating in the \lya forest, and highlight its sensitivity to different X-ray preheating models using numerical simulations.

X-ray preheating affects the \lya forest in two main ways: (1) X-rays wipe out some of the small-scale structures that would be otherwise present, leading to a less transparent IGM at low redshifts. (2) Besides the ``puffing up'' of structures during the early reionization process, X-rays will penetrate deeply into the IGM raising its temperature and therefore setting up an earlier start to cosmic reionization. This early start also eventually leads to a decrease in the cross-correlation of matter and ionized bubble spatial structure. 

Overall, the inclusion of X-ray preheating in the \lya forest leads to a decrease in the memory of reionization, i.e. a decrease on the large-scale enhancement due to the impact of inhomogeneous \ion{H}{I} reionization in the forest, at low redshifts with the total change depending on the specific X-ray preheating prescription. Furthermore, we find that the deviation from a model with no X-ray preheating can almost reach the $10\%$ mark for the 3D flux power spectrum of our Type I boxes at high redshift and for directions perpendicular to the line of sight, which also coincides with the sweet spot for the detection of the memory of reionization in the \lya forest \citep{2021MNRAS.508.1262M}. Likewise, the deviations in the already observed 1D flux power spectrum can lead to a change larger than $6\%$ at high redshifts. However, these quantitative findings are subject to the inability of Type I boxes to reliably sample the linear regime in the redshift range $2 \leq z \leq 4$. Thus, we make a separate analysis with the aim to quantify the changes we should expect due to X-ray preheating in what we suggest should be implemented into future fiducial models of the \lya forest interested in accurate IGM evolution in the post-reionization era.

Regarding the larger Type II boxes, our findings for the flux power spectra are encouraging for the current and next generation of \lya forest surveys. At low redshifts, where the reionization relics are lesser and the forest is subject to other astrophysical systematics like UV clustering \citep{2014PhRvD..89h3010P,2014MNRAS.442..187G,2019MNRAS.487.5346T,2023MNRAS.520..948L} and \ion{He}{II} reionization \citep{2013MNRAS.435.3169C,2017ApJ...841...87L,2020MNRAS.496.4372U}, the long-lasting impact of X-ray preheating results in a decrease in the strength of the memory of reionization. In contrast, at large redshifts, where the impact of reionization is a dominant effect in the \lya forest, X-ray preheating leads to a reinforcement of the signal. This enhancement may imply yet more optimistic outcomes when the \lya forest's ability to discern different X-ray preheating scenarios is forecasted or eventually measured.

During the completion of this work, \cite{2023arXiv230208506M} identified a potential issue in {\tt 21cmFAST} simulations, namely that it underpredicts adiabatic fluctuations by approximately an order of magnitude due to the assumption that $T_k$ is homogeneous at $z = 35$ (see their Figure 13). Although these findings can result in large effects on the 21~cm signal, we do not expect significant changes in the memory of reionization since X-ray preheating eventually takes over.       

Our results underscore the necessity of including not only the memory of reionization in the \lya forest but also the long-lasting effects of X-ray preheating in the fiducial model of \lya analyses. Nevertheless, our findings also raise the same warning for other sensitive probes of the IGM in the post-reionization era. Thus, future work will investigate the impact of X-ray preheating in the imprints of \ion{H}{I} reionization present in 21~cm intensity mapping \citep{2022arXiv221002385L}.

\section*{Acknowledgements}

We thank the anonymous referee for their insightful comments. This work is supported by the National SKA Program of China (Grant No.~2020SKA0110401), the Major Key Project of PCL, NSFC (Grant No.~11821303, 12050410236), and the National Key R\&D Program of China (Grant No.~2018YFA0404502). PMC was partially supported by the Tsinghua Shui Mu Scholarship during the completion of this work. 
We are grateful to Catalina Morales-Gutierrez and Chris Hirata for their useful suggestions and comments. The authors acknowledge the Tsinghua Astrophysics High-Performance Computing platform at Tsinghua University for providing computational and data storage resources that have contributed to the research results reported within this paper. This work made extensive use of the \hyperlink{https://ui.adsabs.harvard.edu}{NASA Astrophysics DataSystem} and the following open-source python libraries: \texttt{matplotlib} \citep{2007CSE.....9...90H}, \texttt{numpy} \citep{2020Natur.585..357H}, and \texttt{scipy} \citep{2020NatMe..17..261V}. 

\section*{Data Availability}

The data underlying this article will be shared on reasonable request to the corresponding authors.



\bibliographystyle{mnras}
\bibliography{x_ray} 




\appendix

\section{Temperature-density relation}
\label{app:hemd}
Here we focus on the effect of X-ray preheating as seen in the temperature-density relation and its evolution.

It is expected that the heated gas from cosmic reionization would eventually recover a tight temperature-density relation $T \propto \Delta^{\gamma-1}$ \citep{1997MNRAS.292...27H,2016MNRAS.456...47M}, where $T$ is the gas temperature and $\Delta = \rho/\bar{\rho}$ is the ratio of the density and mean density. However, the bimodal nature of the temperature-density relation, caused by the way the small-scale structure reionized \citep{2018MNRAS.474.2173H}, requires cosmological timescales to relax the additional injected energy during the reionization process. Thus, the recovery of the temperature-density relation, even for models with X-ray preheating, is delayed until $z \sim 2$\footnote{Naturally \ion{He}{II} reionization will also disturb the temperature-density relation. However, no high-mass resolution analog study has been done to investigate the way the small-scale structure reionizes. Hence, the study of the \emph{memory} of \ion{He}{II} reionization and its impact on the \ion{H}{I} relics will be left to explore in future work.}. Given the strength of X-rays on the transparency of the IGM, we expect that differentiating between different X-ray prescriptions would be difficult. Thus we opt to showcase only the Fiducial model here and compare it with the $\zeta_X3$ model since the latter showcases the largest deviations for late reionization scenarios. 

\begin{figure*}
    \begin{subfigure}{0.33\linewidth}
        \includegraphics[height=1.15\linewidth, angle=-90]{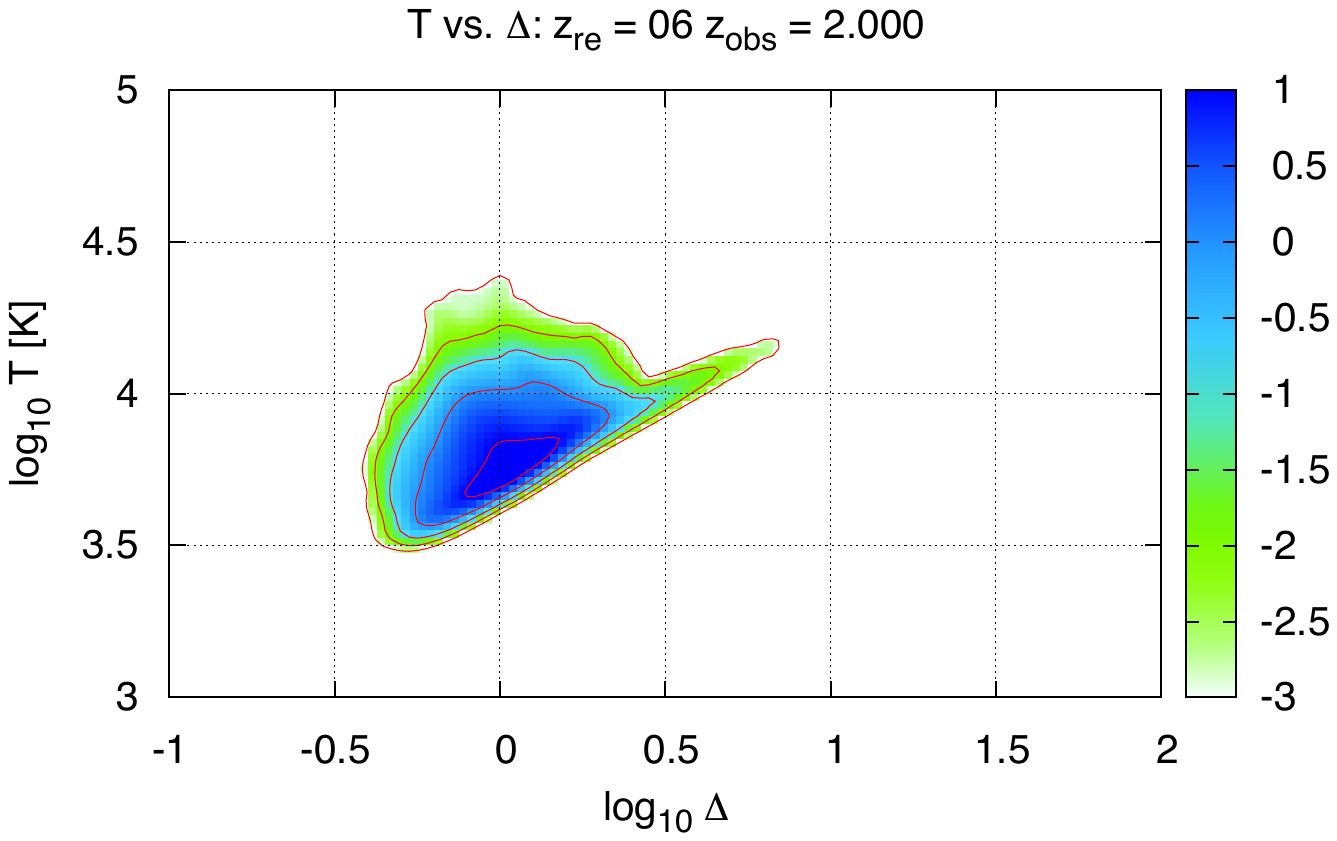}
    \end{subfigure}
    \begin{subfigure}{0.33\linewidth}
        \includegraphics[height=1.15\linewidth, angle=-90]{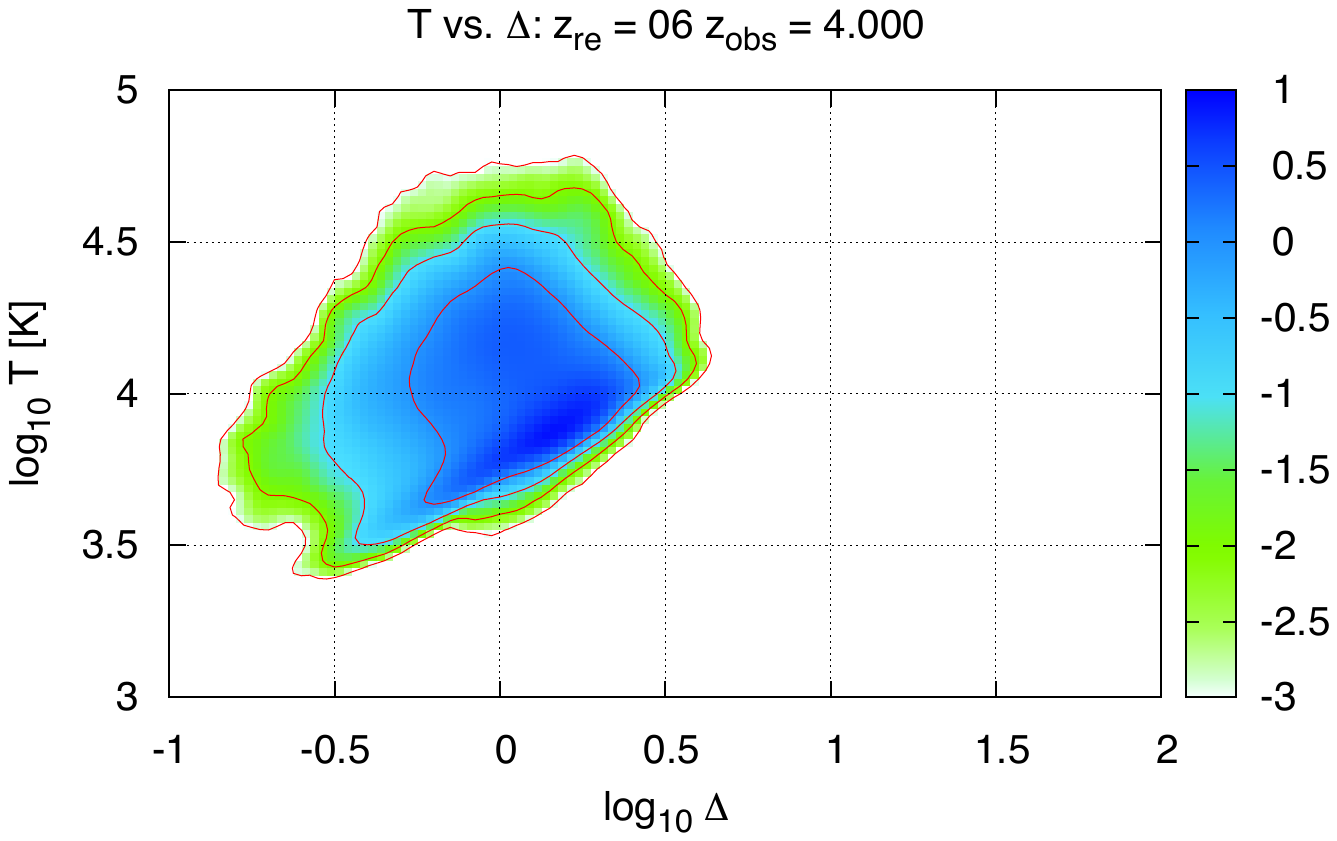}
    \end{subfigure}
    \begin{subfigure}{0.33\linewidth}
        \includegraphics[height=1.15\linewidth, angle=-90]{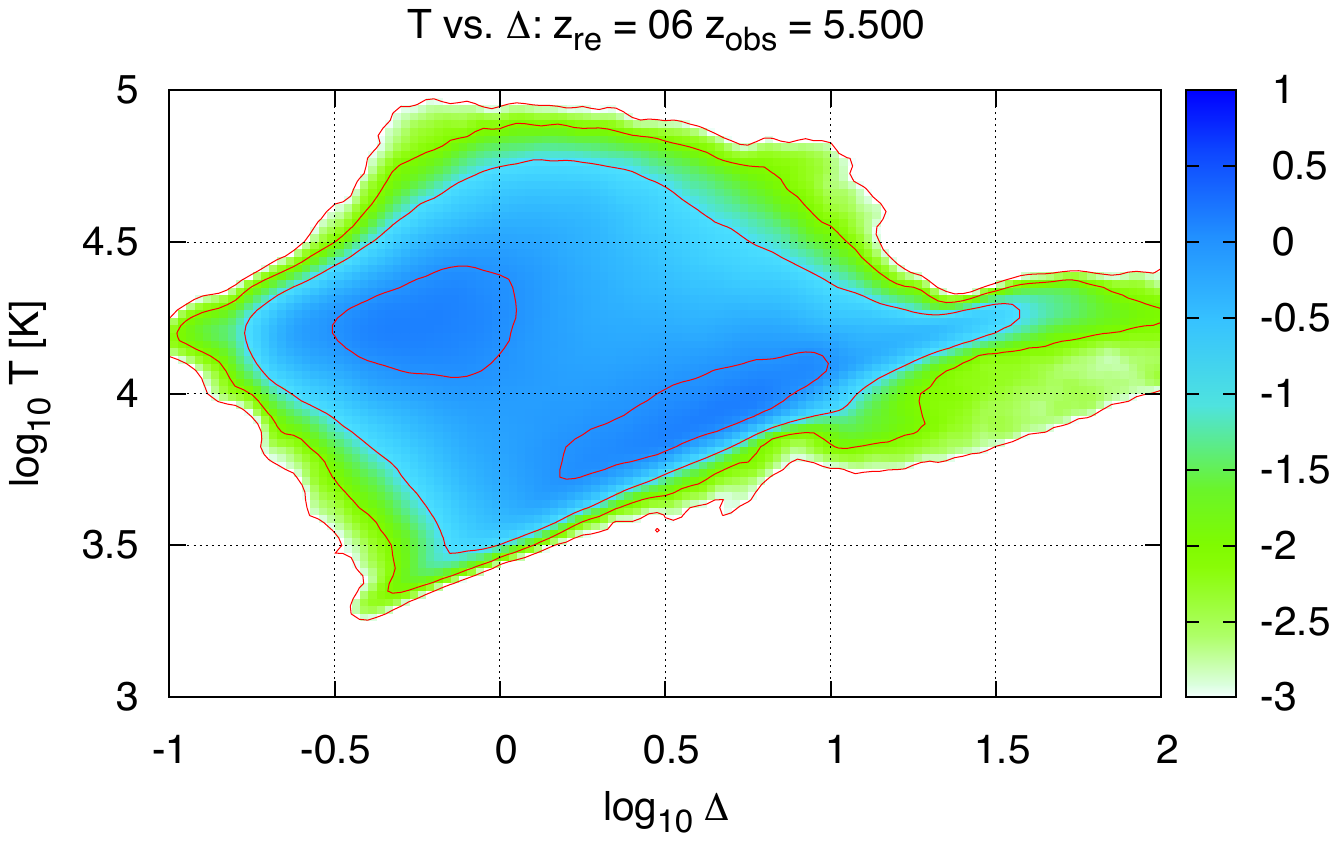}
    \end{subfigure} \\
    \begin{subfigure}{0.33\linewidth}
        \includegraphics[height=1.15\linewidth, angle=-90]{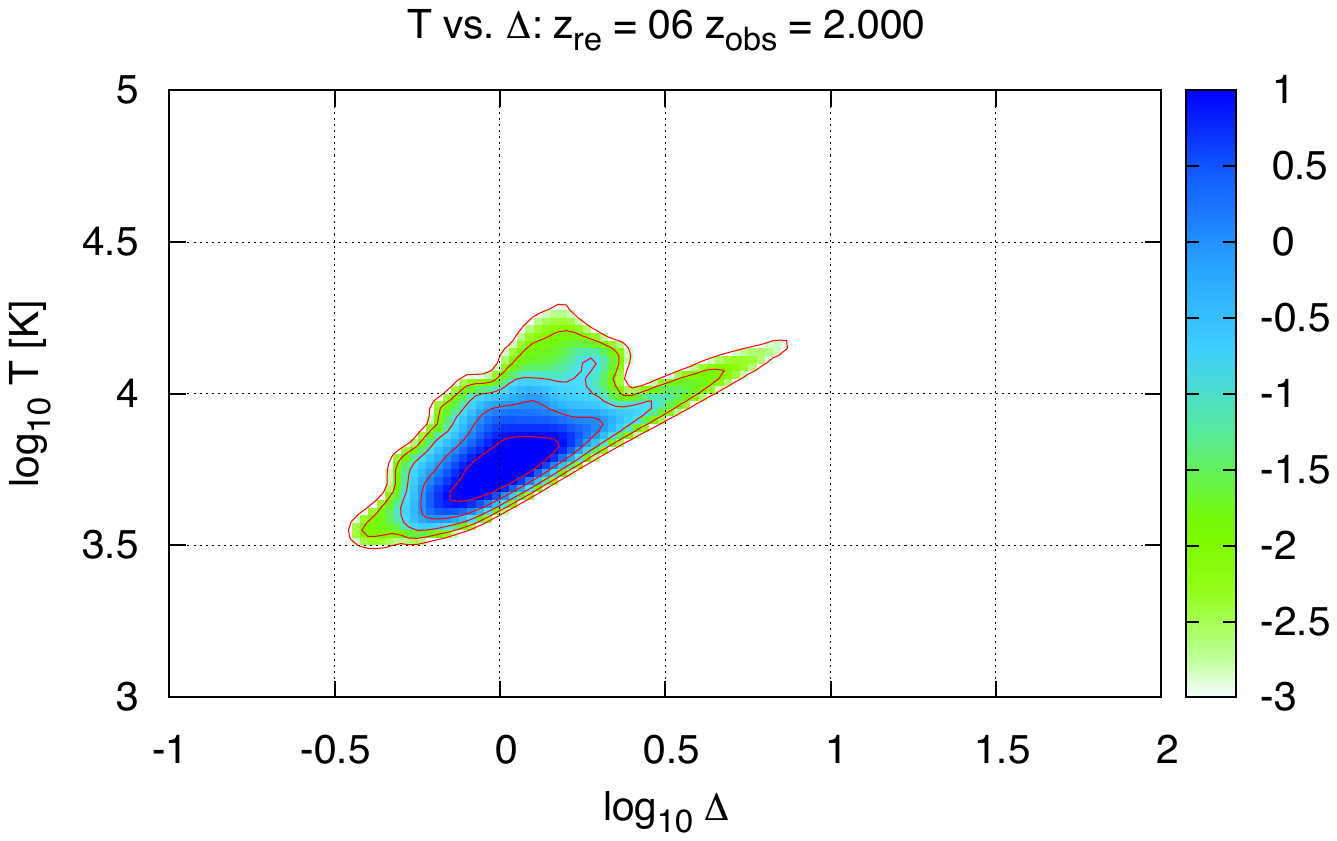}
    \end{subfigure}
    \begin{subfigure}{0.33\linewidth}
        \includegraphics[height=1.15\linewidth, angle=-90]{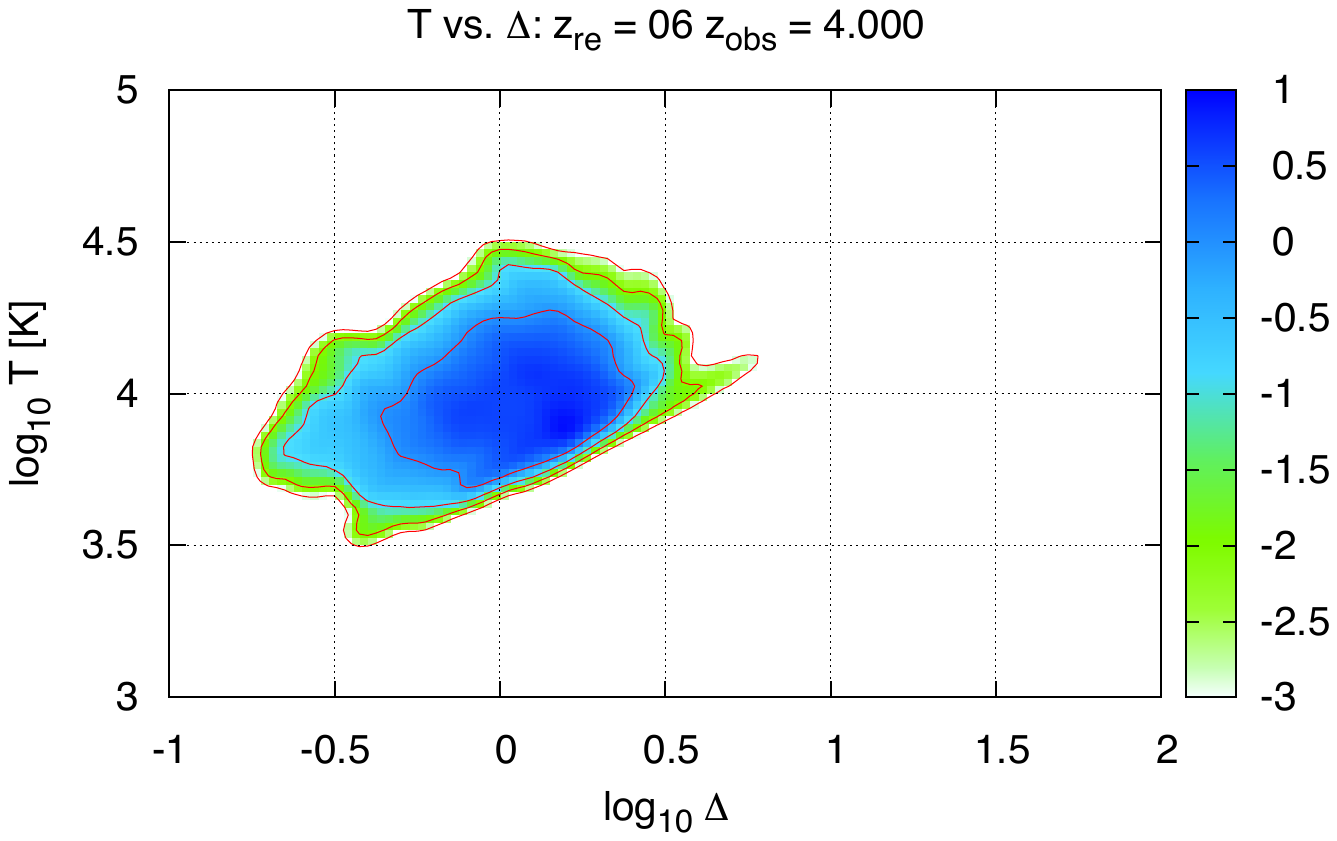}
    \end{subfigure}
    \begin{subfigure}{0.33\linewidth}
        \includegraphics[height=1.15\linewidth, angle=-90]{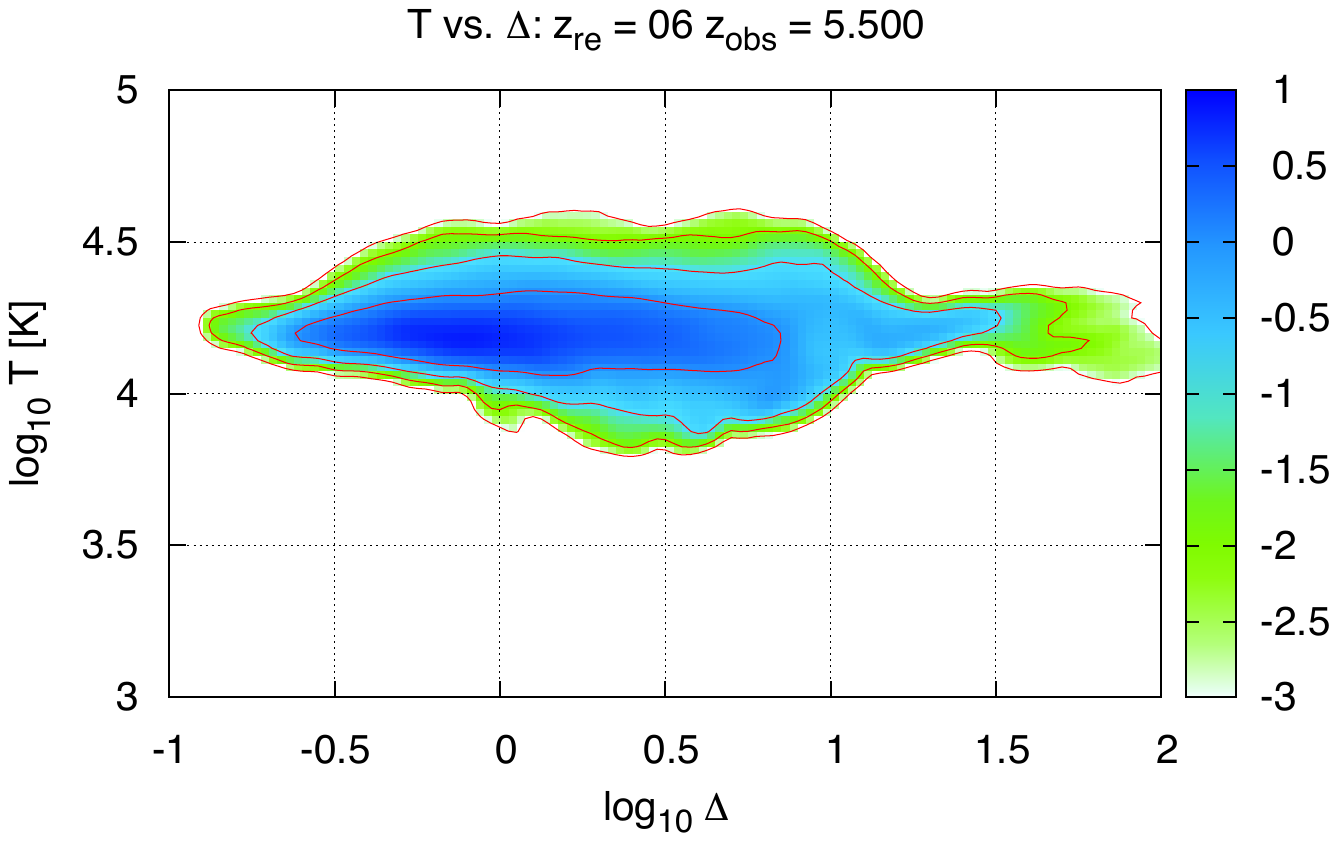}
    \end{subfigure}
    \caption{The evolution of temperature-density relation in the post-reionization era for the latest local reionization redshift ($z_{\rm re} = 6$), which is consistent with recent constraints for the tail end of reionization \citep[see, e.g.][]{2020MNRAS.491.1736K}. The gas particles have been smoothed using a Gaussian density estimator with an FWHM of 0.05 dex on each axis. The color scale shows the mass-weighted distribution of gas in units of 
    $\log_{10}$ probability per $\log_{10}\Delta$ per $\log_{10}T$.
    Contours are shown for every decade in probability density. Shown are the results at the redshift of observation $z_{\rm obs} = 2.0$ (left), $4.0$ (middle), and $5.5$ (right), for different X-ray preheating models --- Fiducial model (i.e.\ no X-ray preheating; top row), and the $\zeta_X3$ model (i.e.\ X-ray preheating is taken from the {\tt 21cmFAST} prescription; bottom row), respectively.}
    \label{fig:hemd-zre6}
\end{figure*}

\begin{figure*}
    \begin{subfigure}{0.33\linewidth}
        \includegraphics[height=1.15\linewidth, angle=-90]{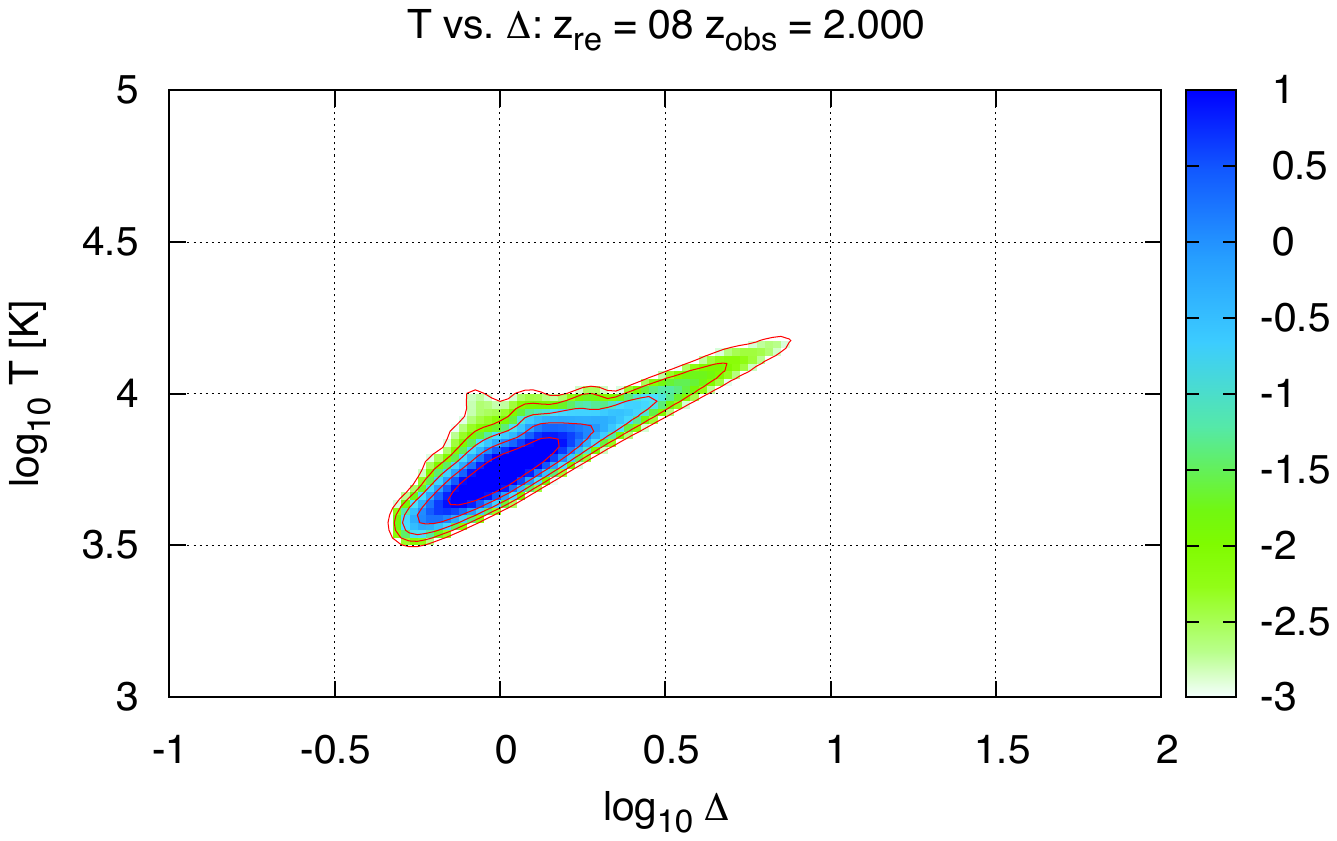}
    \end{subfigure}
    \begin{subfigure}{0.33\linewidth}
        \includegraphics[height=1.15\linewidth, angle=-90]{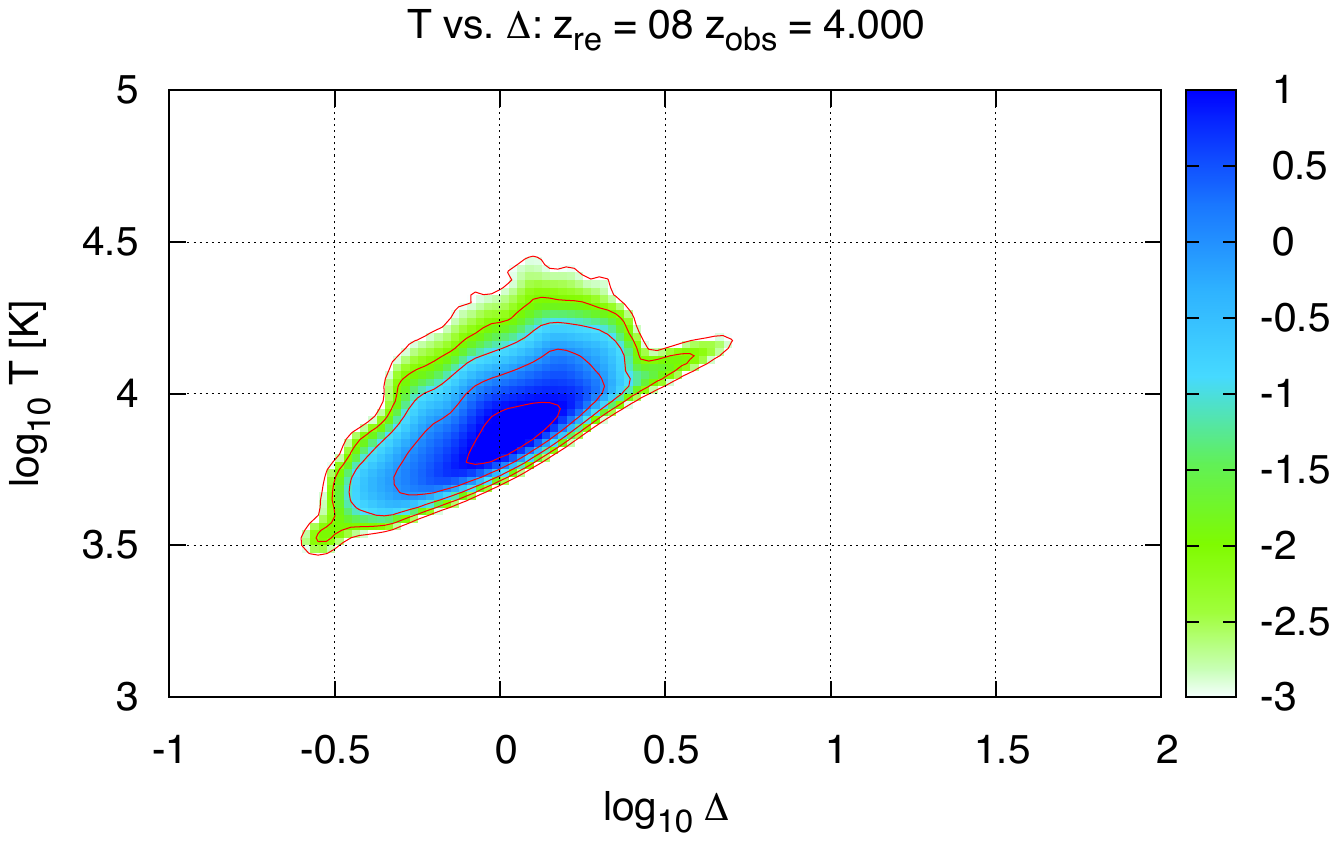}
    \end{subfigure}
    \begin{subfigure}{0.33\linewidth}
        \includegraphics[height=1.15\linewidth, angle=-90]{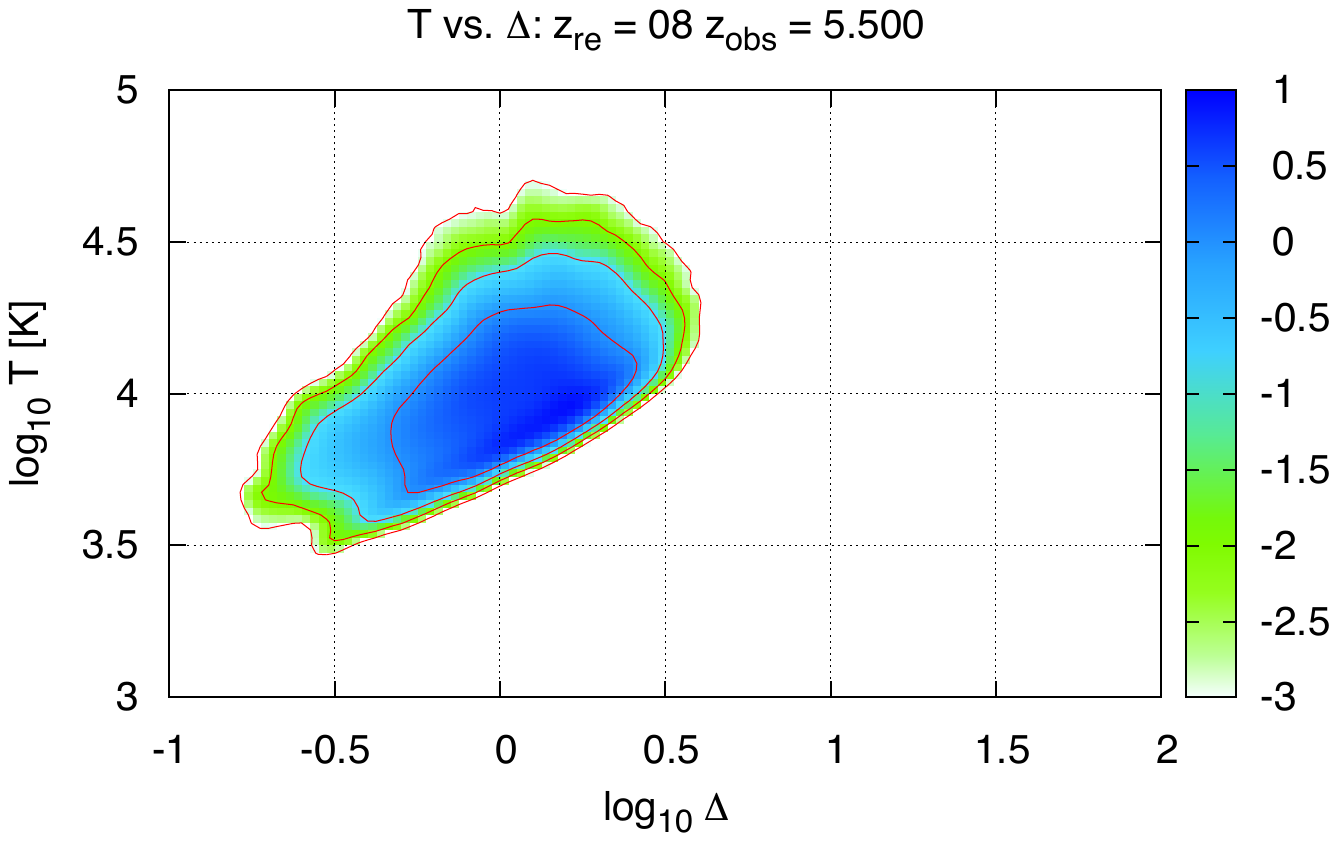}
    \end{subfigure} \\
    \begin{subfigure}{0.33\linewidth}
        \includegraphics[height=1.15\linewidth, angle=-90]{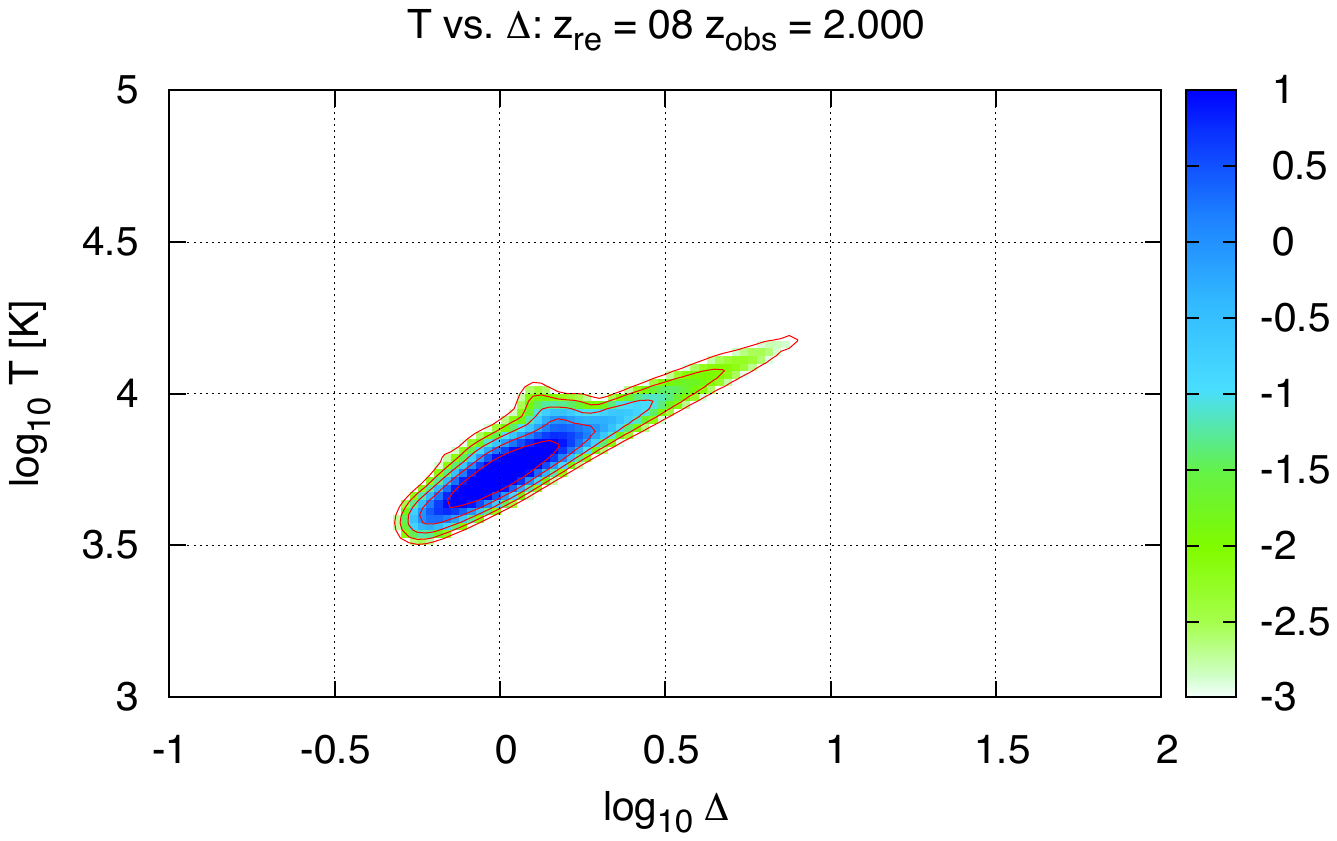}
    \end{subfigure}
    \begin{subfigure}{0.33\linewidth}
        \includegraphics[height=1.15\linewidth, angle=-90]{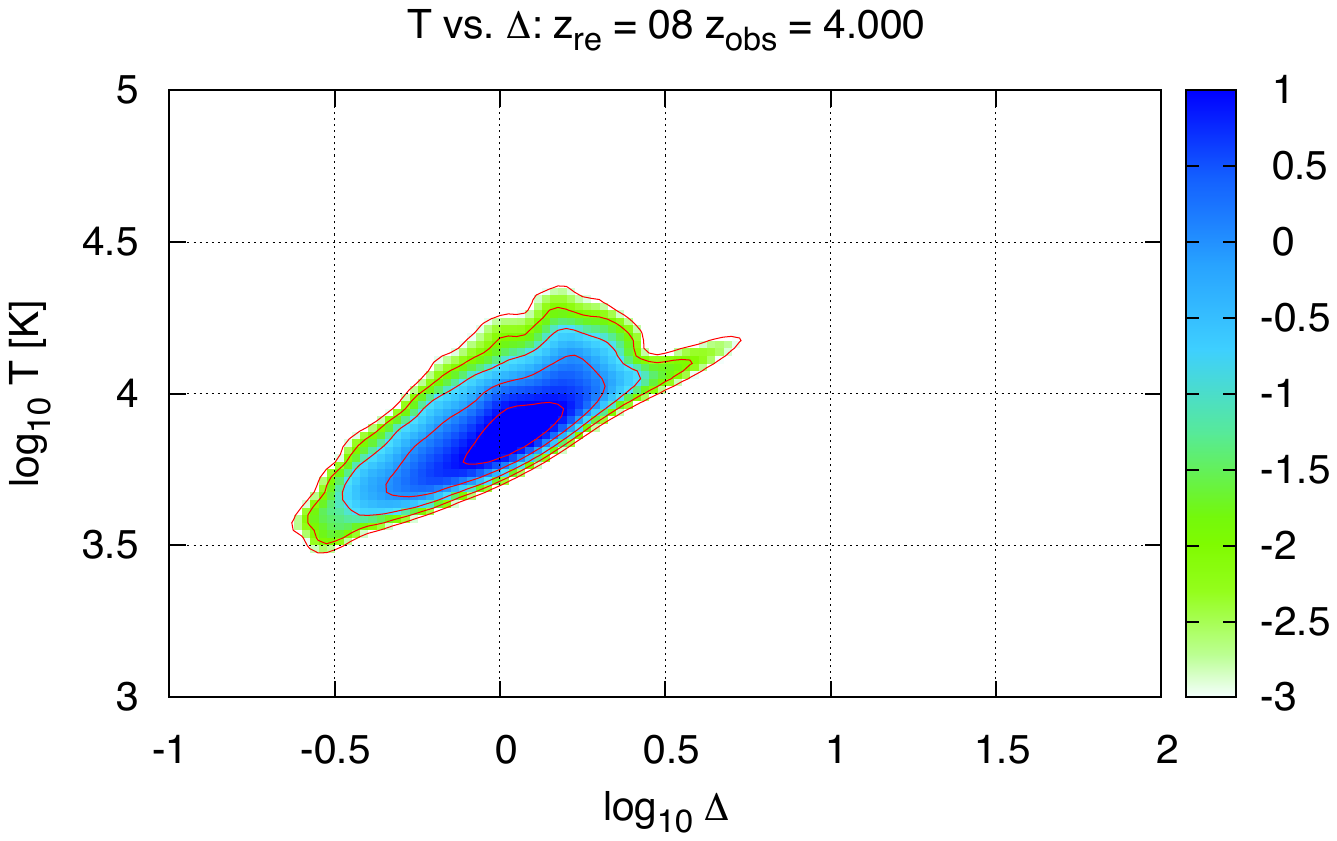}
    \end{subfigure}
    \begin{subfigure}{0.33\linewidth}
        \includegraphics[height=1.15\linewidth, angle=-90]{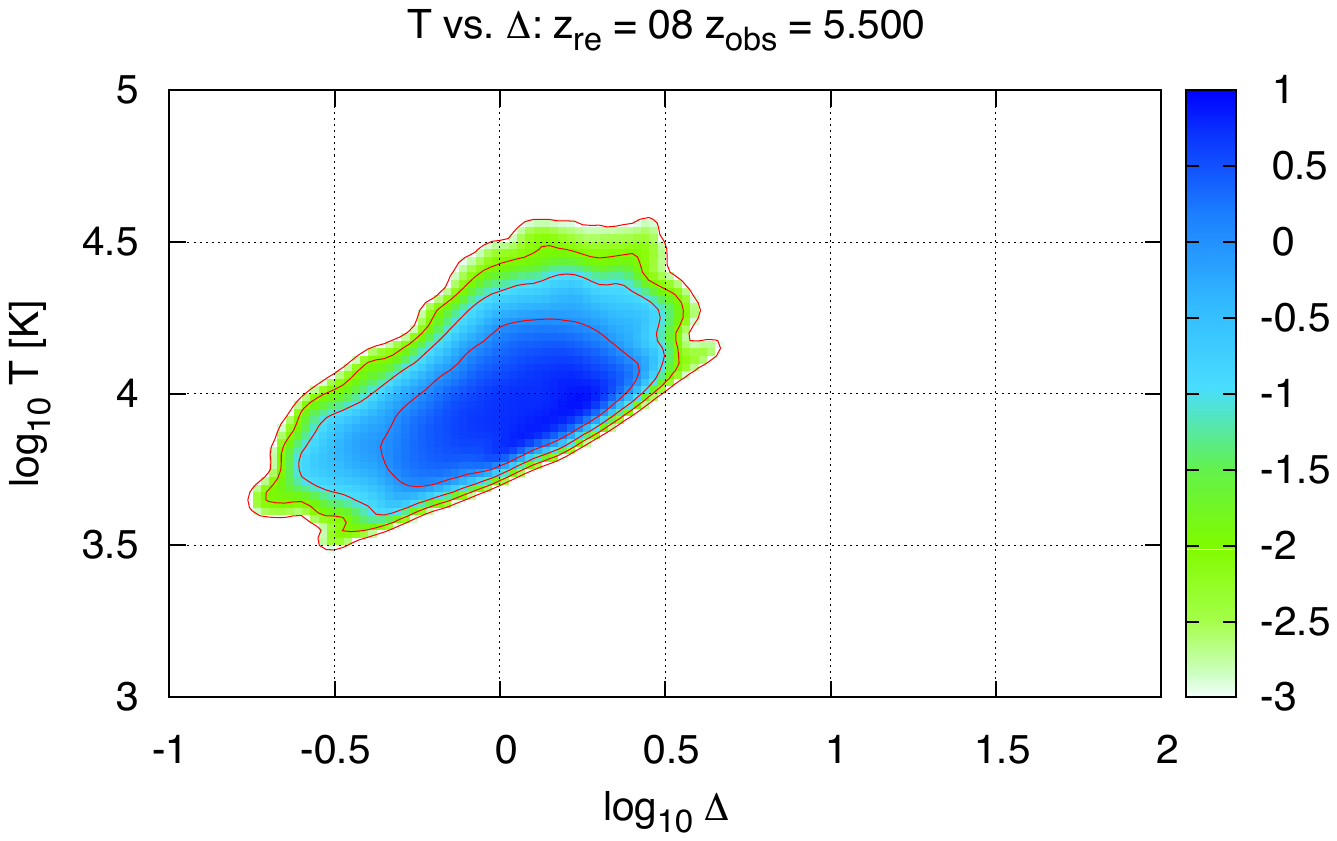}
    \end{subfigure}
    \caption{Same as Figure~\ref{fig:hemd-zre6} but with the redshift of local reionization $z_{\rm re} = 8$ \citep[in line with the constraints from][]{2018arXiv180706209P}.}
    \label{fig:hemd-zre8}
\end{figure*}

Figure~\ref{fig:hemd-zre6} and Figure~\ref{fig:hemd-zre8} illustrate the impact of X-ray preheating in the bi-modal nature of the temperature-density relation for patches of gas that have late local reionization ($z_{\rm re} = 6$) and local reionization consistent with \citealt{2018arXiv180706209P} ($z_{\rm re} = 8$), respectively. In both figures, the Fiducial model (no X-ray heating) leads to a more spread out distribution of the gas, i.e.\ a less tight blue contour, hence a stronger High-Entropy Mean-Density (HEMD; \citealt{2018MNRAS.474.2173H}; \citetalias{2020MNRAS.499.1640M}) phase of the temperature-density relation, than the $\zeta_X3$ model (shown in the bottom panel). As time evolves, the HEMD gas, which originates in the underdense regions that were heated by shocks and UV radiation (and compressed to mean density), traces the evolution of gas in denser regions. Thus, we see a partial recovery of the temperature-density relation at $z_{\rm obs} = 2$ for the late local reionization scenario and a more complete recovery for the Planck local reionization scenario. This recovery is more pronounced in the model with X-rays because some small-scale structures that would otherwise be present were smoothed out. 

\section{IGM tables}
\label{app:I}
We tabulate the relative transparency, as a percentage, of our Type I and Type II boxes with respect to a reference scenario with local reionization occurring at redshift 8 in Tables~\ref{tab:psi} and \ref{tab:psi-II}, respectively.

\begin{table*}
    \centering
    \begin{tabular}{cccccccccc}
    \hline\hline
		Model &  $z_{\rm obs}$ & $b_{\rm \Gamma}$ & \multicolumn{7}{c}{$100 \times \Delta \ln \tau_1$    [$\%$]}  \\
        {} & {} & {} & $z_{\rm re} = 6.0$ & $z_{\rm re} = 7.0$ & $z_{\rm re} = 8.0$ & $z_{\rm re} = 9.0$ & $z_{\rm re} = 10$ & $z_{\rm re} = 11$ & $z_{\rm re} = 12.0$ \\
        \hline
        \multirow{5}{*}{\rotatebox[origin=c]{90}{Fiducial}} & 2.0 & 0.125 & 6.296 $\pm$ 0.392 & 2.443 $\pm$ 0.161 & 0 & -1.308 $\pm$ 0.258 & -1.956 $\pm$ 0.425 & -2.307 $\pm$ 0.565 & -2.518 $\pm$ 0.710 \\
        {} & 2.5 & 0.217 & 7.840 $\pm$ 0.563 & 3.297 $\pm$ 0.151 & 0 & -1.920 $\pm$ 0.212 & -2.764 $\pm$ 0.331 & -3.144 $\pm$ 0.433 & -3.424 $\pm$ 0.546 \\
		{} & 3.0 & 0.346 & 9.655 $\pm$ 0.751 & 4.390 $\pm$ 0.236 & 0 & -2.520 $\pm$ 0.119 & -3.708 $\pm$ 0.177 & -4.234 $\pm$ 0.232 & -4.446 $\pm$ 0.294 \\
        {} & 3.5 & 0.513 & 12.358 $\pm$ 0.880 & 5.771 $\pm$ 0.219 & 0 & -3.801 $\pm$ 0.201 & -5.571 $\pm$ 0.284 & -6.319 $\pm$ 0.265 & -6.700 $\pm$ 0.323 \\
        {} & 4.0 & 0.712 & 18.049 $\pm$ 4.742 & 8.288 $\pm$ 1.158 & 0 & -5.541 $\pm$ 0.836 & -8.438 $\pm$ 1.148 & -9.713 $\pm$ 1.198 & -10.301 $\pm$ 0.921 \\
        \hline
        \multirow{5}{*}{\rotatebox[origin=c]{90}{Hirata-300}} & 2.0 & 0.125 & 4.067 $\pm$ 0.526 & 1.414 $\pm$ 0.194 & -0.412 $\pm$ 0.093 & -1.471 $\pm$ 0.290 & -2.025 $\pm$ 0.439 & -2.339 $\pm$ 0.575 & -2.531 $\pm$ 0.716 \\
        {}       & 2.5 & 0.217 & 5.095 $\pm$ 0.718 & 1.950 $\pm$ 0.237 & -0.515 $\pm$ 0.064 & -2.109 $\pm$ 0.212 & -2.835 $\pm$ 0.324 & -3.174 $\pm$ 0.433 & -3.438 $\pm$ 0.547 \\
		{}       & 3.0 & 0.346 & 6.707 $\pm$ 0.519 & 2.656 $\pm$ 0.173 & -0.703 $\pm$ 0.042 & -2.774 $\pm$ 0.121 & -3.799 $\pm$ 0.174 & -4.272 $\pm$ 0.232 & -4.463 $\pm$ 0.294 \\
        {}       & 3.5 & 0.512 & 10.087 $\pm$ 1.670 & 3.895 $\pm$ 0.548 & -0.851 $\pm$ 0.131 & -4.103 $\pm$ 0.188 & -5.674 $\pm$ 0.279 & -6.358 $\pm$ 0.267 & -6.718 $\pm$ 0.325 \\
        {}       & 4.0 & 0.710 & 17.939 $\pm$ 5.987 & 6.631 $\pm$ 2.035 & -0.885 $\pm$ 0.378 & -5.869 $\pm$ 0.690 & -8.519 $\pm$ 1.082 & -9.728 $\pm$ 1.165 & -10.307 $\pm$ 0.907 \\
        \hline
        \multirow{5}{*}{\rotatebox[origin=c]{90}{Fast-Fid}} & 2.0 & 0.125 & 4.696 $\pm$ 0.507 & 1.682 $\pm$ 0.187 & -0.281 $\pm$ 0.052 & -1.408 $\pm$ 0.276 & -1.989 $\pm$ 0.432 & -2.315 $\pm$ 0.569 & -2.518 $\pm$ 0.710 \\
        {}        & 2.5 & 0.217 & 5.871 $\pm$ 0.626  & 2.299 $\pm$ 0.221 & -0.358 $\pm$ 0.044 & -2.036 $\pm$ 0.211 & -2.798 $\pm$ 0.328 & -3.152 $\pm$ 0.434 & -3.424 $\pm$ 0.546 \\
		{}        & 3.0 & 0.346 & 7.506 $\pm$ 0.499 & 3.090 $\pm$ 0.162 & -0.491 $\pm$ 0.031 & -2.675 $\pm$ 0.121 & -3.752 $\pm$ 0.176 & -4.244 $\pm$ 0.234 & -4.446 $\pm$ 0.294 \\
        {}        & 3.5 & 0.513 & 10.675 $\pm$ 1.464 & 4.351 $\pm$ 0.486 & -0.595 $\pm$ 0.097 & -3.984 $\pm$ 0.190 & -5.620 $\pm$ 0.281 & -6.328 $\pm$ 0.266 & -6.699 $\pm$ 0.322 \\
        {}        & 4.0 & 0.710 & 17.913 $\pm$ 5.680 & 7.014 $\pm$ 1.812 & -0.627 $\pm$ 0.272 & -5.736 $\pm$ 0.739 & -8.475 $\pm$ 1.115 & -9.716 $\pm$ 1.187 & -10.301 $\pm$ 0.918 \\
        \hline
        \multirow{5}{*}{\rotatebox[origin=c]{90}{$\zeta_X1$}} & 2.0 & 0.125 & 5.413 $\pm$ 0.445 & 2.098 $\pm$ 0.150 & -0.077 $\pm$ 0.121 & -1.307 $\pm$ 0.323 & -1.932 $\pm$ 0.468 & -2.279 $\pm$ 0.598 & -2.491 $\pm$ 0.735 \\
        {}        & 2.5 & 0.217 & 6.741 $\pm$ 0.629 & 2.839 $\pm$ 0.156 & -0.098 $\pm$ 0.146 & -1.932 $\pm$ 0.270 & -2.738 $\pm$ 0.379 & -3.115 $\pm$ 0.470 & -3.398 $\pm$ 0.573 \\
		{}        & 3.0 & 0.346 & 8.420 $\pm$ 0.675 & 3.771 $\pm$ 0.323 & -0.132 $\pm$ 0.199 & -2.509 $\pm$ 0.256 & -3.668 $\pm$ 0.260 & -4.194 $\pm$ 0.292 & -4.412 $\pm$ 0.334 \\
        {}        & 3.5 & 0.513 & 11.308 $\pm$ 1.143 & 5.093 $\pm$ 0.325 & -0.143 $\pm$ 0.248 & -3.768 $\pm$ 0.388 & -5.506 $\pm$ 0.423 & -6.260 $\pm$ 0.360 & -6.648 $\pm$ 0.378 \\
        {}        & 4.0 & 0.711 & 17.765 $\pm$ 5.341 & 7.645 $\pm$ 1.437 & -0.136 $\pm$ 0.273 & -5.508 $\pm$ 0.937 & -8.353 $\pm$ 1.252 & -9.649 $\pm$ 1.283 & -10.251 $\pm$ 0.995 \\
        \hline
        \multirow{5}{*}{\rotatebox[origin=c]{90}{$\zeta_X2$}} & 2.0 & 0.125 & 3.836 $\pm$ 0.547 & 1.229 $\pm$ 0.207 & -0.468 $\pm$ 0.160 & -1.474 $\pm$ 0.345 & -2.004 $\pm$ 0.478 & -2.309 $\pm$ 0.605 & -2.503 $\pm$ 0.738 \\
        {}        & 2.5 & 0.217 & 4.806 $\pm$ 0.759 & 1.709 $\pm$ 0.239 & -0.596 $\pm$ 0.147 & -2.135 $\pm$ 0.256 & -2.811 $\pm$ 0.372 & -3.144 $\pm$ 0.468 & -3.409 $\pm$ 0.572 \\
		{}        & 3.0 & 0.346 & 6.414 $\pm$ 0.538 & 2.364 $\pm$ 0.252 & -0.806 $\pm$ 0.153 & -2.766 $\pm$ 0.237 & 3.762 $\pm$ 0.253 & -4.230 $\pm$ 0.290 & -4.426 $\pm$ 0.333 \\
        {}        & 3.5 & 0.512 & 9.872 $\pm$ 1.789 & 3.590 $\pm$ 0.504 & -0.958 $\pm$ 0.152 & -4.076 $\pm$ 0.334 & -5.614 $\pm$ 0.403 & -6.301 $\pm$ 0.354 & -6.664 $\pm$ 0.377 \\
        {}        & 4.0 & 0.709 & 17.962 $\pm$ 6.191 & 6.382 $\pm$ 2.163 & -0.967 $\pm$ 0.377 & -5.823 $\pm$ 0.796 & -8.446 $\pm$ 1.173 & -9.668 $\pm$ 1.254 & -10.257 $\pm$ 0.986 \\
        \hline
        \multirow{5}{*}{\rotatebox[origin=c]{90}{$\zeta_X3$}} & 2.0 & 0.125 & 2.747 $\pm$ 0.692 & 0.559 $\pm$ 0.323 & -0.812 $\pm$ 0.243 & -1.638 $\pm$ 0.385 & -2.080 $\pm$ 0.494 & -2.345 $\pm$ 0.615 & -2.518 $\pm$ 0.743 \\
        {}        & 2.5 & 0.217 & 3.489 $\pm$ 0.981 & 0.831 $\pm$ 0.325 & -1.033 $\pm$ 0.190 & -2.313 $\pm$ 0.270 & -2.891 $\pm$ 0.369 & -3.179 $\pm$ 0.467 & -3.425 $\pm$ 0.574 \\
		{}        & 3.0 & 0.345 & 5.137 $\pm$ 0.748 & 1.287 $\pm$ 0.302 & -1.386 $\pm$ 0.140 & -3.018 $\pm$ 0.231 & -3.862 $\pm$ 0.245 & -4.272 $\pm$ 0.287 & -4.444 $\pm$ 0.332 \\
        {}        & 3.5 & 0.512 & 8.944 $\pm$ 1.821 & 2.480 $\pm$ 0.571 & -1.634 $\pm$ 0.141 & -4.371 $\pm$ 0.313 & -5.727 $\pm$ 0.385 & -6.345 $\pm$ 0.349 & -6.683 $\pm$ 0.376 \\
        {}        & 4.0 & 0.708 & 18.092 $\pm$ 6.376 & 5.459 $\pm$ 2.529 & -1.627 $\pm$ 0.616 & -6.131 $\pm$ 0.668 & -8.543 $\pm$ 1.096 & -9.691 $\pm$ 1.216 & -10.266 $\pm$ 0.972 \\
        \hline
        \multirow{5}{*}{\rotatebox[origin=c]{90}{E1}} & 2.0 & 0.125 & 3.130 $\pm$ 0.656 & 1.635 $\pm$ 0.164 & -0.226 $\pm$ 0.041 & -1.377 $\pm$ 0.275 & -1.973 $\pm$ 0.430 & -2.308 $\pm$ 0.566 & -2.514 $\pm$ 0.709 \\
        {}        & 2.5 & 0.217 & 4.156 $\pm$ 0.987 & 2.238 $\pm$ 0.207 & -0.288 $\pm$ 0.037 & -2.000 $\pm$ 0.216 & -2.781 $\pm$ 0.330 & -3.144 $\pm$ 0.434 & -3.421 $\pm$ 0.546 \\
		{}        & 3.0 & 0.346 & 5.977 $\pm$ 0.966 & 3.020 $\pm$ 0.197 & -0.396 $\pm$ 0.029 & -2.626 $\pm$ 0.123 & -3.729 $\pm$ 0.177 & -4.235 $\pm$ 0.234 & -4.442 $\pm$ 0.295 \\
        {}        & 3.5 & 0.513 & 9.534 $\pm$ 1.022 & 4.273 $\pm$ 0.437 & -0.478 $\pm$ 0.082 & -3.928 $\pm$ 0.199 & -5.595 $\pm$ 0.284 & -6.319 $\pm$ 0.265 & -6.696 $\pm$ 0.322 \\
        {}        & 4.0 & 0.711 & 17.636 $\pm$ 5.155 & 6.940 $\pm$ 1.816 & -0.506 $\pm$ 0.223 & -5.677 $\pm$ 0.776 & -8.453 $\pm$ 1.128 & -9.711 $\pm$ 1.195 & -10.298 $\pm$ 0.920 \\
        \hline
        \multirow{5}{*}{\rotatebox[origin=c]{90}{E2}} & 2.0 & 0.125 & 5.691 $\pm$ 0.439 & 2.120 $\pm$ 0.174 & -0.123 $\pm$ 0.021 & -1.353 $\pm$ 0.271 & -1.968 $\pm$ 0.429 & -2.308 $\pm$ 0.566 & -2.514 $\pm$ 0.709 \\
        {}        & 2.5 & 0.217 & 7.096 $\pm$ 0.548 & 2.871 $\pm$ 0.183 & -0.158 $\pm$ 0.023 & -1.973 $\pm$ 0.216 & -2.776 $\pm$ 0.331 & -3.144 $\pm$ 0.434 & -3.421 $\pm$ 0.546 \\
		{}        & 3.0 & 0.346 & 8.810 $\pm$ 0.611 & 3.809 $\pm$ 0.177 & -0.219 $\pm$ 0.025 & -2.589 $\pm$ 0.123 & -3.724 $\pm$ 0.176 & -4.234 $\pm$ 0.234 & -4.443 $\pm$ 0.295 \\
        {}        & 3.5 & 0.513 & 11.638 $\pm$ 1.067 & 5.134 $\pm$ 0.334 & -0.267 $\pm$ 0.057 & -3.883 $\pm$ 0.203 & -5.588 $\pm$ 0.282 & -6.319 $\pm$ 0.265 & -6.695 $\pm$ 0.322 \\
        {}        & 4.0 & 0.711 & 17.860 $\pm$ 5.074 & 7.693 $\pm$ 1.420 & -0.287 $\pm$ 0.135 & -5.629 $\pm$ 0.801 & -8.449 $\pm$ 1.131 & -9.711 $\pm$ 1.195 & -10.299 $\pm$ 0.920 \\
        \hline
        \multirow{5}{*}{\rotatebox[origin=c]{90}{E3}} & 2.0 & 0.125 & 6.066 $\pm$ 0.406  & 2.340 $\pm$ 0.166 & -0.028 $\pm$ 0.006 & -1.312 $\pm$ 0.262 & -1.951 $\pm$ 0.425 & -2.300 $\pm$ 0.563 & -2.512 $\pm$ 0.707 \\
        {}        & 2.5 & 0.217 & 7.550 $\pm$ 0.553 & 3.159 $\pm$ 0.161 & -0.035 $\pm$ 0.010 & -1.926 $\pm$ 0.215 & -2.759 $\pm$ 0.332 & -3.137 $\pm$ 0.434 & -3.418 $\pm$ 0.546 \\
		{}        & 3.0 & 0.346 & 9.313 $\pm$ 0.694 & 4.182 $\pm$ 0.203 & -0.049 $\pm$ 0.013 & -2.527 $\pm$ 0.123 & -3.702 $\pm$ 0.178 & -4.226 $\pm$ 0.234 & -4.439 $\pm$ 0.295 \\
        {}        & 3.5 & 0.513 & 12.031 $\pm$ 0.939 & 5.562 $\pm$ 0.257 & -0.059 $\pm$ 0.019 & -3.807 $\pm$ 0.206 & -5.564 $\pm$ 0.286 & -6.311 $\pm$ 0.265 & -6.692 $\pm$ 0.322 \\
        {}        & 4.0 & 0.712 & 17.879 $\pm$ 4.859 & 8.084 $\pm$ 1.253 & -0.064 $\pm$ 0.033 & -5.544 $\pm$ 0.836 & -8.428 $\pm$ 1.153 & -9.709 $\pm$ 1.203 & -10.297 $\pm$ 0.921 \\
        \hline\hline
    \end{tabular}
    \caption{Relative transparency of the Type I simulations defined using the Fiducial model, which has no X-ray preheating, reionizing at redshift 8 as a reference as follows: $\psi \times 10^2 = 10^2 \times \Delta \ln \tau_1  = 10^2 \times \ln [\tau_1 (z_{\rm obs}, z_{\rm re}) / \overline{\tau}_1 (z_{\rm obs}, 8)$].}
    \label{tab:psi}
\end{table*}

\begin{table*}
    \centering
    \begin{tabular}{cccccccccc}
    \hline\hline
		Model &  $z_{\rm obs}$ & $b_{\rm \Gamma}$ & \multicolumn{7}{c}{$100 \times \Delta \ln \tau_1$    [$\%$]}  \\
        {} & {} & {} & $z_{\rm re} = 6.0$ & $z_{\rm re} = 7.0$ & $z_{\rm re} = 8.0$ & $z_{\rm re} = 9.0$ & $z_{\rm re} = 10$ & $z_{\rm re} = 11$ & $z_{\rm re} = 12.0$ \\
        \hline
        \multirow{5}{*}{\rotatebox[origin=c]{90}{Fiducial}} & 2.0 & 0.094 & 5.954 $\pm$ 0.439 & 2.075 $\pm$ 0.235 & 0 & -0.802 $\pm$ 0.250 & -1.074 $\pm$ 0.422 & -1.171 $\pm$ 0.596 & -1.202 $\pm$ 0.686 \\
        {} & 2.5 & 0.160 & 6.643 $\pm$ 0.369 & 2.284 $\pm$ 0.184 & 0 & -0.415 $\pm$ 0.433 & 0.067 $\pm$ 0.762 & 0.651 $\pm$ 1.045  & 1.093 $\pm$ 1.203 \\
		{} & 3.0 & 0.253 & 9.061 $\pm$ 0.800 & 3.379 $\pm$ 0.191 & 0 & -0.623 $\pm$ 0.480 & 0.083 $\pm$ 0.876 & 1.113 $\pm$ 1.272 & 2.052 $\pm$ 1.449 \\
        {} & 3.5 & 0.370 & 13.586 $\pm$ 0.607 & 5.259 $\pm$ 0.354 & 0 & -1.888 $\pm$ 0.523 & -1.645 $\pm$ 1.022 & -0.618 $\pm$ 1.399 & 0.410 $\pm$ 1.661 \\
        {} & 4.0 & 0.509 & 21.596 $\pm$ 0.695 & 8.684 $\pm$ 0.264 & 0 & -3.962 $\pm$ 0.329 & -4.860 $\pm$ 0.773 & -4.433 $\pm$ 1.107 & -3.564 $\pm$ 1.329 \\
        \hline
        \multirow{5}{*}{\rotatebox[origin=c]{90}{Fast-Fid}} & 2.0 & 0.094 & 4.781 $\pm$ 0.502 & 1.585 $\pm$ 0.230 & -0.161 $\pm$ 0.060 & -0.845 $\pm$ 0.282 & -1.090 $\pm$ 0.435 & -1.177 $\pm$ 0.601 & -1.204 $\pm$ 0.688 \\
        {} & 2.5 & 0.159 & 5.769 $\pm$ 0.339 & 1.972 $\pm$ 0.111 & -0.030 $\pm$ 0.048 & -0.387 $\pm$ 0.451 & 0.089 $\pm$ 0.767 & 0.661 $\pm$ 1.039 & 1.099 $\pm$ 1.196 \\
		{} & 3.0 & 0.251 & 8.784 $\pm$ 0.713 & 3.216 $\pm$ 0.262 & 0.138 $\pm$ 0.033 & -0.483 $\pm$ 0.467 & 0.168 $\pm$ 0.850 & 1.159 $\pm$ 1.237 & 2.075 $\pm$ 1.413 \\
        {} & 3.5 & 0.368 & 14.592 $\pm$ 0.364 & 5.603 $\pm$ 0.258 & 0.341 $\pm$ 0.068 & -1.620 $\pm$ 0.507 & -1.481 $\pm$ 0.976 & -0.524 $\pm$ 1.346 & 0.466 $\pm$ 1.607 \\
        {} & 4.0 & 0.506 & 25.061 $\pm$ 1.448 & 9.752 $\pm$ 0.561 & 0.611 $\pm$ 0.206 & -3.554 $\pm$ 0.324 & -4.606 $\pm$ 0.723 & -4.281 $\pm$ 1.049 & -3.466 $\pm$ 1.278 \\
        \hline\hline
    \end{tabular}
    \caption{Same as Table~\ref{tab:psi} but for the select Type II simulations. We restrict the number of models considered in the Type II analysis due to the limitation in computational resources and the non-trivial task of distinguishing X-ray prescriptions.}
    \label{tab:psi-II}
\end{table*}     

We highlight that there can be significant deviations between the results for $\Delta \ln \tau_1$ for the Type I and II boxes. For instance, at $z_{\rm re} = 9$ and $z_{\rm obs} = 2.5$ the Fast-Fid and Fiducial models report an approximately 80 percent difference (ignoring the error bars). In contrast, at $z_{\rm re} = 7$  and $z_{\rm obs} = 2.5$ the Fast-Fid and Fiducial models have a deviation of 6 and 15 percent, respectively. The discrepancy between boxes arises from the difficulty the Type I box has to capture the forest due to poor sampling. However, we acknowledge that convergence for the transparency result has not been achieved even with the larger Type II boxes. 

Our Type II boxes correspond to \cite{2018MNRAS.474.2173H} II-C boxes and have a missing variance of 1.32 at $z = 2.5$ (see their \S5.3), which is an improvement from the 2.67 missing variance of the Type I boxes that have a factor of 9 smaller volume. Nevertheless, convergence in $\Delta \ln \tau_1$ for even the II-F box ($L = 2551$ kpc, $216$ times the Type I volume, and a missing variance of 0.77) used in \citetalias{2019MNRAS.487.1047M} has not been achieved. The necessary computational resources to achieve satisfactory convergence in $\Delta \ln \tau_1$ are too restrictive, particularly given the significant number of simulations needed to compute our variance, which is why we have chosen to use our Type II boxes for this analysis.


\bsp	
\label{lastpage}
\end{document}